\documentstyle[aps,prb,epsf]{revtex} 

\begin{document} 

\draft

\title{Excitation spectra for Andreev billiards of Box and Disk geometries}

\author{J. Cserti, A. Bodor, J. Koltai and G. Vattay\\ 
E{\"o}tv{\"o}s University, Department of Physics of 
Complex Systems, \\
H-1117 Budapest, P\'azm\'any P{\'e}ter s{\'e}t\'any 1/A, Hungary}


\maketitle

\begin{abstract}
We study Andreev billiards of box and disk geometries by matching
the wave functions at the interface of the normal and the superconducting
region using the exact solutions of the Bogoliubov-de Gennes equation.
The mismatch in the Fermi wavenumbers and the effective 
masses of the normal system and the superconductor, as well as 
the tunnel barrier at the interface are taken into account. 
A Weyl formula (for the smooth part of the counting
function of the energy levels) is derived. 
The exact quantum mechanical calculations show 
equally spaced singularities in the density of states. 
Based on the Bohr-Sommerfeld quantization rule a semiclassical theory 
is proposed to understand these singularities. 
For disk geometries two kinds of states can be distinguished:
states either contribute through whispering gallery modes or are 
Andreev states strongly coupled to the superconductor.
Controlled by two relevant material parameters, three kinds of energy spectra
exist in disk geometry.
The first is dominated by Andreev reflections, 
the second, by normal reflections in an  annular disk geometry.
In the third case the coherence length is much larger than the radius of 
the superconducting region, and the spectrum is identical to that of
a full disk geometry.
\end{abstract}



\section{Introduction} \label{sec:intro}

Recent technological advances in manufacturing almost ballistic 
semiconductors of mesoscopic size (two dimensional electron gas, 2DEG) 
coupled to a superconductor has initiated a growing interest in 
considering the phase-coherent transport and the excitation spectrum 
in hybrid superconducting nanostructures.
In these systems the electron can coherently evolve into a hole (and 
vice versa) at the interface between the semiconductor and 
the superconductor. 
This mechanism had been discovered by Andreev\cite{Andreev}. 
Classically, the particle momentum is conserved and the hole is 
reflected back in the same direction as the incoming electron. 
This scattering process is known as retro-reflection or Andreev reflection.  
The overview of the recent progress in this field can be found in several
works\cite{Colin1,Carlo-konyv,Schon1,Carlo1}.
The Andreev scattering resulting in a discrete spectrum 
of single-particle excitations of a layer of 
normal metal in contact with superconductors on both sides 
was first discussed by Andreev\cite{Andreev}.
Based on the Bogoliubov-de Gennes equation\cite{BdG-eq} (BdG) 
the excitation spectrum (Andreev states) of a normal metal (N) 
attached to a superconductor (S) was first considered 
by P.\ G.\ de~Gennes and D.\ Saint-James \cite{deGennes-Saint-James}. 
A ballistic normal metal weakly coupled to a superconductor is commonly
called an Andreev billiard. Such systems have been extensively studied
over the past ten years
\cite{Kosztin,Melsen,Melsen-2,Altland,Lesovik,Nazarov,Heny,Richter1}.  
The bound states were studied e.g. for SNS 
junctions\cite{Kulik_Bardeen,Carlo_2_3,Bagwell}. 
A semi-infinite N region in contact with a semi-infinite S region
in a strong magnetic field was investigated 
by Hoppe et al.\ \cite{Ulrich}

The excitation spectrum of Andreev billiards depends on whether the N
region is classically chaotic or 
integrable\cite{Melsen,Nazarov,Heny,Richter1}. 
It has been shown that a gap opens in the spectrum for chaotic billiards, 
while for integrable billiards the spectrum is most likely to 
be gapless\cite{Melsen}.
In these works the spectrum was calculated only close to the Fermi level. 
However, the spectrum shows interesting features throughout 
the entire energy range below the superconducting gap.  
The excitation spectrum is determined in this paper 
for two specific Andreev billiards, namely the NS box and disk systems. 
In NS box systems a rectangular N region is in contact with the
superconductor, while the NS disk system consists of a circular S region
surrounded concentrically by a circular N region. 
Both systems are integrable and, in accordance with earlier findings, 
the density of states (DOS) is gapless. We shall show, however, that for
higher energies the DOS has singularities located at equal distances 
from each other.

One of the central issues in this paper is the investigation 
of these singularities in the DOS. 
We calculate the spectra of these systems exactly 
by solving the BdG equations, taking into account the non-perfect interface
(mismatch in the Fermi wave numbers and the effective masses of the normal 
metal and the superconductor, and tunnel barrier at the interface).
For a narrow NS junction it is justified
\cite{Colin1,Carlo-konyv,Carlo1} 
that the pairing potential can be approximated by a step-like function 
(zero in the N region and some constant value in the S region).   
Note that the two systems have the common feature that
the BdG equation is separable and one of the separated wave function sets 
(`transverse modes') is the same in the N and the S region.
The exact quantization condition can be very simply expressed in terms
of a phase $\Phi_m(\varepsilon)$ which is shown to be related to the
classical action of an electron moving inside the N region between 
two successive bounces at the NS interface (see Eq.\ (\ref{Phi_m})).

Several authors \cite{Melsen,Melsen-2,Nazarov,Heny,Richter1} have 
already derived the Bohr-Sommerfeld approximation for the density of
states. The DOS can be written in terms of the probability 
distribution $P(s)$ of the classical path length $s$
between two subsequent bounces of the electron at the NS interface.
Starting from our exact quantization condition expressed in terms of the
phase $\Phi_m(\varepsilon)$  we re-derive the commonly used Bohr-Sommerfeld 
approximation\cite{Melsen,Heny,Richter1} of the DOS in the case of NS box
and disk systems. Moreover, we give an analytical expression for the 
probability $P(s)$ in terms of the phase $\Phi_m(\varepsilon)$. 

It is shown that, in the framework of Bohr-Sommerfeld quantization, 
the singularity of the DOS is a direct consequence of the singular 
behavior of $P(s)$. Using the expression for $P(s)$ we can
reproduce the counting function very accurately in the entire range of 
the spectrum and locate the singularities of the DOS for NS box and 
disk systems. The probability $P(s)$ depends on the geometry of 
the N region and the location of the NS interface. 
We shall show that from a semiclassical point of view some special 
trajectories of the electron play crucial roles in the singular
character of the DOS.  
Based on our semiclassical analysis a simple formula is given for the
location of these singularities. This formula gives an excellent agreement 
with our numerically exact calculation of the spectrum. 
Furthermore, it predicts very accurately  
the number of the singularities and their location in the system studied 
by de Gennes et al.\ \cite{deGennes-Saint-James} 
We propose an explanation, based on our theory about the singularities,  
for the pronounced peaks found by Ihra et al.\ \cite{Richter1} 
(for NS systems) in their numerical calculations of the DOS.
We believe that the singularities found by 
Lodder and Nazarov\cite{Nazarov} (for Andreev billiards), and 
\v{S}ipr et al.\ \cite{Gyorffy} (for SNS systems) are related to
some special classical trajectories of the electron for which the probability
$P(s)$ is singular.  

The other issue studied in this paper is the Weyl formula for our NS 
systems. In the quantization of a normal billiard, the counting 
function $N(E)$ gives the number of levels whose energy is
less than or equal to $E$. The smooth part of 
the counting function $N(E)$ is given by the Weyl
formula\cite{Weyl,Brack}, which has already been calculated 
for billiards of various shapes. 
We present a new Weyl formula for our NS box and disk systems. 
As expected, the counting function obtained from the exact
(numerical) calculations `oscillates' around the curve given by 
our Weyl formula.

Our exact quantum mechanical results reveal a more complex 
energy level structure for NS disk systems than for box systems.  
Semiclassically, two kinds of modes exist in the disk geometry. 
First, the electron can hit the superconductor undergoing an Andreev 
reflection (hereafter called Andreev states), and second,  
the electron trajectory may not reach the superconductor (hereafter
called whispering gallery modes). The two modes play crucial roles in the
energy levels of the system. Contrary to NS box systems where the DOS
is proportional to the energy close the Fermi level, in the case of NS
disk systems it is constant due to the whispering gallery modes. 
In studying the spectrum of the disk geometry, 
we have been motivated by Bruder and Imry's recent work\cite{Bruder}.
Based on whispering gallery modes, they proposed a physical picture to
interpret the significant paramagnetic effect observed in recent 
experiments\cite{Visani}. 
Thus, a careful analysis of Andereev and whispering gallery 
states as presented in this paper and the inclusion of the flux induced 
phase may shed further light on the issue raised 
by Bruder and Imry\cite{Bruder}.   
Our work on this problem is in progress. 

We also found that, depending on the parameters of the NS disk system, the
spectra can belong to either of three classes: (a) `mixed phase', in which 
Andreev states coexist with whispering gallery modes and the 
energy dependent coherence length $\xi(\varepsilon)$ in the 
superconductor is much less than the radius $R_{\rm S}$ of the S region, 
while the DOS is singular; 
(b) the opposite case, i.e.\ $\xi(\varepsilon) \gg R_{\rm S}$ 
when the spectrum is identical to that of a 
normal billiard whose radius is that of the outer circle; 
(c) when $k_{\rm F}\xi(0)$ is order of one (here $k_{\rm F}$ is the Fermi
wave number) or the NS interface is not perfect, and so the energy spectrum 
corresponds to that of an annular billiard (the circular S region is
cut out). 
It turns out that there are two relevant parameters to
make this classification, $k_{\rm F}\xi(\varepsilon)$ and 
$k_{\rm F}R_{\rm S}$.
Using these parameters we sketch a so-called `phase diagram' for the 
three above classes.  
It was reported in the works on the paramagnetic 
effect\cite{Visani} that the coherence length exceeds the
size of the superconducting region.  Thus, according to our phase diagram 
it is possible that the whole NS disk system behaves as a full normal
disk. Then a larger paramagnetic effect can be expected 
than that predicted by Bruder and Imry who included only the
whispering gallery states. However, further work is necessary
to clarify this scenario. 

The text is organized as follows. In Sec.\ \ref{seq:secular-eq} 
a quantization condition (secular equation) is derived from 
the matching conditions of the wave functions at the interface of 
the NS systems. Owing to the symmetry of the BdG equation, the secular
equation can be expressed in a very compact form valid both for NS box
and disk systems. In Sec.\ \ref{sec:Weyl} the Weyl formula is derived 
for NS box and disk systems and compared with the exact results.
Our theory for the singularities in the DOS is presented in Sec.\ \ref{sing}.
We sketch the so-called `phase diagram' 
for the NS disk systems in Sec.\ \ref{fazis-diag}, and 
the conclusions are given in Sec.\ \ref{veg}.

\section{Secular Equation for Box and Disk} \label{seq:secular-eq} 

\begin{figure}[hbt]
{\centerline{\leavevmode \epsfxsize=7cm \epsffile{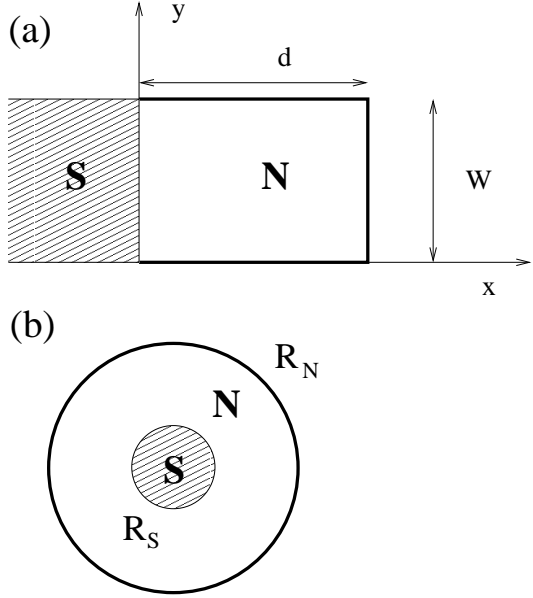 }}}
\caption{The normal system (N) is in contact with the superconducting 
region (S). The two  geometrical arrangements are (a) box
geometry of side lengths $d$ and $W^\prime$ and (b) disk geometry 
(two concentric circles of radii $R_S$ and $R_N$). 
For the box, the generalized coordinates are $(x_1,x_2)=(x,y)$, while 
for the disk $(x_1,x_2)=(r,\varphi)$ centered at the origin of the
superconducting circle. 
\label{abra-gen-box-disk}}
\end{figure}

In this section we derive a secular equation determining the energy levels
for box and disk geometries shown 
in Fig.\ \ref{abra-gen-box-disk}. It is possible to treat both
problems  in a common framework by introducing the generalized coordinates
\begin{equation}
(x_1,x_2)=\left\{ \begin{array}{l} 
(x,y), \,\,\,\,\,  \mbox{for box}, \\[1ex]
(r,\varphi), \,\,\,\,\,  \mbox{for disk.} 
\end{array} \right. 
\end{equation}
The normal region is in contact with a superconducting region at 
$x_1=x_{\rm{NS}}$ (where $x_{\rm{NS}}=0$ for box and 
$x_{\rm{NS}}=R_{\rm S}$ for disk). 

At the interface $x_1=x_{\rm{NS}}$ the tunnel barrier is modeled 
by a Dirac delta potential $V(x_1,x_2)=U_0\, \delta(x_1-x_{\rm{NS}})$,  
as in Ref.~\onlinecite{BTK}. 
The superconducting pairing potential is a constant $\Delta_0$ in the 
S region and zero in the N region. 
The self-consistency of the pairing potential is not taken into
account just like in Ref.~\onlinecite{BTK}. 
For a planar geometry the difference between the self-consistent gap and
the step function is quite small as it was shown by
McMillan and Kieselmann\cite{McMillan}. 
Similar results has been found by
Plehn et al.\ in Ref.~\onlinecite{Plehn} for superconducting multilayers. 
For disk geometry the step
function approximation is still reliable provided $R_{\rm S}$ is larger
then the coherence length.   
However, the Fermi energies 
(i.e.\ the energy differences between the band edges of 
the N/S region and the chemical potential\cite{Kato1})
and the effective
masses in the N and S regions are assumed to be different, 
i.e.\ $E_{\rm{F}}^{\rm{(N)}}\neq E_{\rm{F}}^{\rm{(S)}}$ 
and $m_{\rm{N}}\neq m_{\rm{S}}$.

The NS system is described by the BdG equation:
\begin{equation}
\left(\begin{array}{cc}
H_0  & \Delta \\ 
\Delta^* & -H_0 
\end{array}   
 \right)
\Psi = \varepsilon \,  \Psi,
\end{equation}
where $\Psi$ is a two-component wave function, 
$H_0 = {\bf p}^2/2m  - \mu$, and the Fermi energy is  
$\mu = E_{\rm F}^{(N)}$ in the N region and  
$\mu = E_{\rm F}^{(S)}$ in the S region.
The energy levels of the Andreev states are 
the positive eigenvalues $\varepsilon$ of the Bogoliubov-de Gennes (BdG) 
equation\cite{BdG-eq}. In what follows, we consider the energy spectrum
below the superconducting gap, $\varepsilon < \Delta_0$. 
The two-component wave functions $\Psi$ in the N and S 
regions can be chosen in the form
\begin{eqnarray}
\Psi^{\rm{(N)}}_m (x_1,x_2) & = & \left( \begin{array}{c}
 a_m^{(e)} \varphi_m^{(N,e)}(x_1) \\
 a_m^{(h)} \varphi_m^{(N,h)}(x_1) 
\end{array}  \right) \chi_m(x_2), 
\label{hullfv-N} \\[1ex]
 \Psi^{\rm{(S)}}_m (x_1,x_2) & = & 
\left[
c_m^{(e)} \left( \begin{array}{c}\gamma_e \\ 1 \end{array}\right)
\varphi_m^{(S,e)}(x_1)
+ c_m^{(h)}\left( \begin{array}{c}\gamma_h \\ 1 \end{array}\right)
\varphi_m^{(S,h)}(x_1) 
\right] \chi_m(x_2),
\label{hullfv-S}
\end{eqnarray}
where the 'transverse' wave function of the $m$th mode is 
\begin{equation}
\chi_m(x_2)=\left\{ \begin{array}{l} 
\sqrt{2/W} \sin(m\pi y/W), \,\,\,\,\,  \mbox{for box}, \\[1ex]
e^{im\varphi}, \,\,\,\,\,  \mbox{for disk.} 
\end{array} \right. 
\end{equation}
The 'transverse' wave functions 
in the N and S regions are the same, and it is possible to separate the
variables $x_1,x_2$ in the BdG equation, i.e.\  
$\chi_m(x_2)$ depends only on the coordinate $x_2$. 

The wave function $\varphi_m^{(N,e)}(x_1)$ is the electron-like component
of the eigenspinor of the BdG equation in the N region. It satisfies the
one-dimensional Schr{\"o}dinger equation obtained by separating 
$\chi_m(x_2)$ in the BdG equation:
\begin{equation}
\varphi_m^{(N,e)}(x_1)=\left\{ \begin{array}{l} 
\sin k^{(e)}_m (d-x), \,\,\,\,\,  \mbox{for box}, \\[1ex]
J_m(k_e r) - 
\frac{J_m(k_e R_{\rm{N}})}{Y_m(k_e R_{\rm{N}})}
Y_m(k_e r), \,\,\,\,\,  \mbox{for disk,} 
\end{array} \right. 
\label{phi_N_e}
\label{hullfv-box-disk}
\end{equation}
where the energy dependence of the wave number $k^{(e)}_m(\varepsilon)$
for box and $k_e (\varepsilon)$ for disk are given by 
\begin{eqnarray}
k^{(e)}_m(\varepsilon) & = &  k_{\rm{F}}^{\rm{(N)}}
\sqrt{1+\varepsilon/E_{\rm{F}}^{\rm{(N)}}
-{(m \pi/k_{\rm{F}}^{\rm{(N)}}W)}^2}, \,\,\,\,\,  \mbox{for box}, \\[1ex] 
k_e (\varepsilon)& = & k_{\rm{F}}^{\rm{(N)}}
\sqrt{1+ \varepsilon/E_{\rm{F}}^{\rm{(N)}}}, \,\,\,\,\,
\mbox{for disk,}
\end{eqnarray} 
$k_{\rm{F}}^{\rm{(N)}}$ and $E_{\rm{F}}^{\rm{(N)}}$ are 
the Fermi wave number and the Fermi energy in the N region, respectively, 
while $J_m(x)$ and $Y_m(x)$ are the Bessel and Neumann functions of 
order $m$.
As we shall see below it is convenient to use  real wave functions  
for $\varphi_m^{(N,e)}(x_1)$.  
The wave functions have been chosen such that they satisfy 
the Dirichlet boundary conditions at $x=d$ for box and 
$r=R_{\rm{N}}$ for disk.
Due to the symmetry of the BdG equation the hole-like component of the
eigenspinor of the BdG equation in the N region is
\begin{equation}
\varphi_m^{(N,h)}(\varepsilon,x_1)=\varphi_m^{(N,e)}(-\varepsilon,x_1),
\label{symmetry-N}
\end{equation}
where the explicit dependence of the energy $\varepsilon$ 
in the wave functions is emphasized for clarity.

The wave function in the S region is 
\begin{equation}
\varphi_m^{(S,e)}(\varepsilon,x_1)=
\left\{ \begin{array}{l} 
\exp(-iq^{(e)}_m x), \,\,\,\,\,  \mbox{for box}, \\[1ex]
J_m(q_e r), \,\,\,\,\,  \mbox{for disk,} 
\end{array} \right. 
\end{equation} 
where 
\begin{eqnarray}
q^{(e)}_m(\varepsilon) & = &  
k_{\rm{F}}^{\rm{(S)}}
\sqrt{1+ \eta -{(m \pi/k_{\rm{F}}^{\rm{(S)}}W)}^2}, \,\,\,\,\,  
\mbox{for box}, \\[1ex] 
q_e (\varepsilon)& = & k_{\rm{F}}^{\rm{(S)}}\sqrt{1+ \eta}, \,\,\,\,\,
\mbox{for disk,} \label{qe-disk}
\end{eqnarray} 
$k_{\rm{F}}^{\rm{(S)}}$ and $E_{\rm{F}}^{\rm{(S)}}$ are 
the Fermi wave number and the Fermi energy in the S region,
respectively, 
and 
\begin{equation}
\eta =  \sqrt{\varepsilon^2-\Delta_0^2}/E_{\rm{F}}^{\rm{(S)}}.
\end{equation}
Note that for box $\varphi_m^{(S,e)}(x) \rightarrow 0$ as 
$x\rightarrow -\infty$ satisfying the boundary condition at $-\infty$. 
On the other hand, only the Bessel function can be chosen for disks, since
the Neumann function is singular at the origin.
Again, owing to the symmetry of the BdG equation, the hole-like component 
of the BdG spinor in the S region is 
\begin{equation}
\varphi_m^{(S,h)}(\varepsilon,x_1)
={[\varphi_m^{(S,e)}(-\varepsilon,x_1)]}^*.
\label{symmetry-S}
\end{equation}

Finally, in Eq.\ (\ref{hullfv-S}) 
\begin{eqnarray}
\gamma_{e} & = & \Delta_0/(\varepsilon - 
\sqrt{\varepsilon^2 -\Delta_0^2}), \\[1ex]
\gamma_h & = & \gamma_e^*.
\end{eqnarray}

The coefficients $a_m^{(e)}, a_m^{(h)}, c_m^{(e)}, c_m^{(h)}$ 
in Eqs.\ (\ref{hullfv-N})-(\ref{hullfv-S}) 
are determined from the boundary conditions at the interface of the NS
system (see, e.g., Ref.~\onlinecite{Kato1}):
\begin{eqnarray}
\Psi^{\rm{(N)}}_m 
\rule[-1.6ex]{.2pt}{4ex}\;\raisebox{-1.5ex}{$\scriptstyle x_1=x_{\rm{NS}}$}
& = &  \Psi^{\rm{(S)}}_m 
\rule[-1.6ex]{.2pt}{4ex}\;\raisebox{-1.5ex}{$\scriptstyle
x_1=x_{\rm{NS}}$}, 
\nonumber \\
\frac{d}{dx_1} \left[\Psi^{\rm{(N)}}_m 
- \frac{m_{\rm{N}}}{m_{\rm{S}}}\Psi^{\rm{(S)}}_m \right]
\rule[-1.6ex]{.2pt}{4ex}\;\raisebox{-1.5ex}{$\scriptstyle x_1=x_{\rm{NS}}$}  
& = & \frac{2m_{\rm{N}}}{\hbar^2} U_0 \Psi^{\rm{(S)}}_m 
\rule[-1.6ex]{.2pt}{4ex}\;\raisebox{-1.5ex}{$\scriptstyle
x_1=x_{\rm{NS}}$}.
\label{matching}
\end{eqnarray}
The matching conditions yield the following 
secular equation for the eigenvalues $\varepsilon$  of the NS system 
for fixed mode index $m$:
\begin{equation}
D^{\rm{(NS)}}_m(\varepsilon) =
\left| \begin{array}{cccc}
 \varphi_m^{(N,e)} & 0 & 
\gamma_e \varphi_m^{(S,e)} & \gamma_h \varphi_m^{(S,h)}  \\
0 & \varphi_m^{(N,h)}& \varphi_m^{(S,e)} & \varphi_m^{(S,h)} \\
{\left[\varphi_m^{(N,e)}\right]}^\prime & 0 & 
\gamma_e \left(Z\varphi_m^{(S,e)} +\frac{m_{\rm{N}}}{m_{\rm{S}}} 
{\left[\varphi_m^{(S,e)}\right]}^\prime \right) 
& \gamma_h \left( Z\varphi_m^{(S,h)} + \frac{m_{\rm{N}}}{m_{\rm{S}}}
{\left[\varphi_m^{(S,h)}\right]}^\prime \right) \\
0 & {\left[\varphi_m^{(N,h)}\right]}^\prime &
Z\varphi_m^{(S,e)} + \frac{m_{\rm{N}}}{m_{\rm{S}}} 
{\left[\varphi_m^{(S,e)}\right]}^\prime  & 
Z\varphi_m^{(S,h)} + \frac{m_{\rm{N}}}{m_{\rm{S}}} 
{\left[\varphi_m^{(S,h)}\right]}^\prime
\end{array}  \right| =0,
\end{equation}
where $Z= \left(2m_{\rm{N}}/\hbar^2\right) \, U_0$ is the normalized barrier
strength, and the prime stands for the derivative with respect to $x_1$.
All the functions are evaluated at $x_1 = x_{\rm{NS}}$.

Using the fact that the wave functions $\varphi_m^{(N,e)}$ given 
in Eq.\ (\ref{phi_N_e}) are real functions and 
the symmetry relations between the electron-like and hole-like
component of the BdG eigenspinor given 
by Eqs.\ (\ref{symmetry-N}) and (\ref{symmetry-S}), the above determinant
can be simplified.  One can show that the secular equation 
reduces to  
\begin{equation} 
\rm{Im} \left \{\gamma_e D^{\rm{(e)}}_m(\varepsilon) 
D^{\rm{(h)}}_m(\varepsilon) \right \}=0,
\label{DNS}
\end{equation}
where 
\begin{equation}
D^{\rm{(e)}}_m(\varepsilon) =
\left| \begin{array}{cc}
 \varphi_m^{(N,e)} & \varphi_m^{(S,e)}  \\
{\left[\varphi_m^{(N,e)}\right]}^\prime & 
Z\varphi_m^{(S,e)} + \frac{m_{\rm{N}}}{m_{\rm{S}}} 
{\left[\varphi_m^{(S,e)}\right]}^\prime
\end{array} \right| 
\label{De}
\end{equation}
is a 2x2 determinant, and $D^{\rm{(h)}}_m(\varepsilon) =
{\left[D^{\rm{(e)}}_m(-\varepsilon)\right]}^* $. 
The energy levels of the NS systems can be found by solving the 
secular equation (\ref{DNS}) for $\varepsilon$ at a given quantum number $m$.  
The secular equation (\ref{DNS}) is exact in the sense that the usual
Andreev approximation is not assumed\cite{Colin1}. 
The Andreev approximation is valid only 
for $\Delta_0/E_{\rm F} \ll 1$ and quasi-particles whose
incident/reflected directions are approximately 
perpendicular to the interface\cite{Colin1}. 

\subsection{Secular equation for box}

To find the energy levels for a box we shall give an explicit form of 
the secular equation (\ref{DNS}). Inserting the wave functions given 
in Eq.\ (\ref{hullfv-box-disk}) into  Eq.\ (\ref{De}) we obtain
\begin{equation}
D^{\rm{(e)}}_m(\varepsilon) =
 \sin k_m^{(e)} d \left(Z-i\frac{m_{\rm{N}}}{m_{\rm{S}}} q_m^{(e)} 
+  k_m^{(e)} \cot  k_m^{(e)} d \right). 
\label{De-box}
\end{equation} 
Hence, the secular equation given in Eq.\ (\ref{DNS}) becomes
 \begin{equation}
\rm{Im} \left \{\gamma_e \left(Z-i\frac{m_{\rm{N}}}{m_{\rm{S}}} q_m^{(e)} 
+  k_m^{(e)} \cot  k_m^{(e)} d \right)
\left(Z+i\frac{m_{\rm{N}}}{m_{\rm{S}}} q_m^{(h)} 
+  k_m^{(h)} \cot  k_m^{(h)} d \right)\right \}=0,
\label{seq-box}
\end{equation} 
where $k_m^{(h)}(\varepsilon)=k_m^{(e)}(-\varepsilon)$ and 
$q_m^{(h)}(\varepsilon)={\left[q_m^{(e)}(-\varepsilon)\right]}^*$ are 
the wave numbers of the hole-like quasi-particles 
in the N and S regions, respectively. 
For a given quantum number $m$ the energy levels can be
found by solving the secular equation.   
Note that the case $\sin k_m^{(e)} d=0$ corresponds 
to the secular equation of the entire box with  Dirichlet boundary conditions.
Thus, it cannot be zero for the NS system.      
Semiclassically, small wave numbers $k_m^{(e)}$ and $k_m^{(h)}$ in the N
region correspond to quasi-particles incident at
grazing angles on the NS interface.   
In this case the Andreev approximation is not valid\cite{Colin1}. 
However, the secular equation (\ref{seq-box}) is {\it still} exact for 
the energy levels for box geometry.

\subsection{Secular equation for disk}

We now derive the explicit form of the secular equation
for disk geometry. 
 Inserting the wave functions given 
in Eq.\ (\ref{hullfv-box-disk}) into  Eq.\ (\ref{De}) we obtain
\begin{equation}
D^{\rm{(e)}}_m(\varepsilon) =
\left| \begin{array}{cc}
 J_m(k_e R_{\rm S})
-\frac{J_m(k_e R_{\rm N})}{ Y_m( k_{e}R_{\rm N})} 
Y_m( k_{e}R_{\rm S}) & J_m( q_{e}R_{\rm S}) \\[1ex]
k_e \left[ J_m^\prime(k_e R_{\rm S}) 
-\frac{J_m(k_e R_{\rm N})}{ Y_m( k_{e}R_{\rm N})} 
Y_m^\prime(k_e R_{\rm S}) \right] & 
Z J_m( q_{e}R_{\rm S}) + \frac{m_{\rm{N}}}{m_{\rm{S}}} q_e 
 J_m^\prime(q_e R_{\rm S})
\end{array} \right|, 
\label{De-disk}
\end{equation} 
where the primes denote the derivatives of the Bessel functions 
with respect to their arguments. 
For a given angular momentum quantum number $m$ 
the energy levels are the solutions of the secular equation 
given in Eq.\ (\ref{DNS}).

\section{Weyl formula for NS systems} \label{sec:Weyl}

For normal systems the counting function $N(E)$ is   
the number of states whose energy is less than or equal to $E$. 
The derivative of the counting function with respect to the energy 
is the density of states. 
The smooth part of the counting function $\tilde{N}(E)$ for a cavity 
was first derived by Weyl \cite{Weyl} 
(for more details see Ref.~\onlinecite{Brack}).  The leading term of 
$\tilde{N}(E)$ is given by the integral of 
$\Theta(E-H_{\rm{cl}}({\bf p},{\bf r}))$ over the phase space
divided by $h^2$, where $\Theta$ is the Heaviside function and 
$H_{\rm{cl}}$ is the Hamiltonian of the corresponding classical system.  
In two dimensions, and excluding the factor 2 for the spin, the smooth 
part of counting function $\tilde{N}(E)$ is given by 
\begin{equation}
\tilde{N}(E) = \frac{1}{h^2} \,  \int_{H_{\rm{cl}}({\bf p},{\bf r}) \le E} 
\, d^2p \, d^2 r \, 
\Theta(E-H_{\rm{cl}}({\bf p},{\bf r})).
\label{Weyl-normal}
\end{equation} 
For Andreev states of energy $\varepsilon$  less than 
$\Delta_0$ it is not possible to define the classical Hamiltonian 
$H_{\rm{cl}}$, therefore the smooth part of the counting function  
$\tilde{N}(\varepsilon)$ cannot be derived
from Eq.\ (\ref{Weyl-normal}).  
As an alternative method for calculating the DOS one may start 
from the secular equation\cite{Doron1,Uzy}. 
In this section we derive a Weyl formula for the NS systems shown 
in Fig. \ref{abra-gen-box-disk} using the secular equation (\ref{DNS}).
  
For a given quantum number $m$ let us introduce 
the eigenphase $\Phi_m(\varepsilon)$ for the NS system: 
\begin{equation}  
{\left[D^{\rm{(e)}}_m(\varepsilon)\right]}^* / D^{\rm{(e)}}_m(\varepsilon)
= e^{i \Phi_m(\varepsilon)}. 
\label{S_NS}
\end{equation}
Here we have assumed that $D^{\rm{(e)}}_m(\varepsilon) \neq 0$, 
and in the following it will also be supposed that 
$D^{\rm{(h)}}_m(\varepsilon) \neq 0$.
In Sec.\ \ref{fazis-diag} the case  
$D^{\rm{(e)}}_m(\varepsilon)D^{\rm{(h)}}_m(\varepsilon) = 0$ 
will be discussed.  
It is also assumed that the eigenphases $\Phi_m(\varepsilon)$ are 
all monotonic and increasing with  $\varepsilon$.
The secular equation (\ref{DNS}) can be further simplified:
\begin{equation}
\Phi_m(\varepsilon)-\Phi_m(-\varepsilon) 
-2 \arccos \frac{\varepsilon}{\Delta_0} 
= 2n\pi,   
\label{Phi_m}
\end{equation}
where $n=0,\pm 1, \pm 2, \cdots$ .
Equation (\ref{Phi_m}) is a very simple quantization condition 
for the NS system. 
The solutions of this equation give the energy spectrum 
$\varepsilon_{mn}$ below the gap. 
The above-given quantization condition is a convenient 
starting point to calculate the DOS and the smooth part of the 
counting function. However, for numerical purposes Eq.\ (\ref{DNS}) 
may be more suitable. Using (\ref{Phi_m}) 
the density of states   
$\rho(\varepsilon) = 
\sum_{m=1}^{M(\varepsilon)}\sum_{n=-\infty}^{\infty} \, 
\delta(\varepsilon -\varepsilon_{mn})$  
for the NS system  becomes
\begin{equation}
\rho(\varepsilon) = \sum_{m=1}^{M(\varepsilon)} \, 
 \left| \frac{d F_m(\varepsilon)}{d \varepsilon} \right| 
\sum_{n=-\infty}^{\infty} \, \delta(F_m(\varepsilon) -n),
\label{rho-1}
\end{equation}
where $2\pi F_m(\varepsilon) = 
\Phi_m(\varepsilon)-\Phi_m(-\varepsilon) 
-2 \arccos (\varepsilon/\Delta_0)$, 
and $M(\varepsilon)$ is the number of 'transverse modes' depending 
on $\varepsilon$.
Since the $\Phi_m(\varepsilon)$ are all monotonic and increasing 
with $\varepsilon$, the derivatives $d F_m(\varepsilon)/d \varepsilon$  
are positive, and so the modulus sign is superfluous in (\ref{rho-1}).
Applying the Poisson summation formula\cite{Berry-1,Brack} 
to the second sum one finds
\begin{equation}
\rho(\varepsilon) =  \sum_{m=1}^{M(\varepsilon)} \,  
\frac{d F_m(\varepsilon)}{d \varepsilon}  
\left[1+ 2\sum_{k=1}^{\infty} \, 
\cos (2\pi k F_m(\varepsilon) )\right ].
\label{rho-2}
\end{equation}
The DOS can be separated into two parts: 
the smooth part, i.e.\ the $k=0$ term, and the oscillating part, which
comes from the terms  $k \ge 1$ in Eq.\ (\ref{rho-2}).  
Here we consider only the smooth part. Then the Weyl formula, i.e.\ 
the smooth part of the counting function  for NS system can be obtained
from  
$\tilde{N}(\varepsilon) = 
\int_0^\varepsilon \, 
\rho(\varepsilon^\prime) \, d \varepsilon^\prime$, and one
finds
\begin{equation}
\tilde{N}(\varepsilon) = \frac{1}{2\pi} \, 
\sum_{m=1}^{M(\varepsilon)} 
\left[ \Phi_m(\varepsilon)-\Phi_m(-\varepsilon) \right]
+M(\varepsilon)\left(\frac{1}{2}-\frac{1}{\pi} \, 
\arccos \frac{\varepsilon}{\Delta_0} \right). 
\label{Weyl}
\end{equation}
This is our Weyl formula for the NS systems. 
A similar treatment has been mentioned 
by Schomerus and Beenakker\cite{Heny} in deriving the 
Bohr-Sommerfeld approximation of the DOS for Andreev's billiards. 
The number of transverse modes $M(\varepsilon)$ is a discontinuous function 
of $\varepsilon$, therefore in Eq.\ (\ref{Weyl}) the sum has to be replaced
by an integral to get the smoothed version of the counting
function\cite{Uzy}. One can see that the Weyl formula is 
expressed in terms of the eigenphases $\Phi_m(\varepsilon)$ defined 
in Eq.\ (\ref{S_NS}). Note that our Weyl formula is only
valid for those systems in which a common set of the `transverse wave' 
functions can be separated from the wave functions of the N and S regions. 
Lifting this condition is a possible extension of this problem.

\subsection{Weyl formula for box (no mismatch)}
\label{box-Weyl}

We now consider the Weyl formula for a perfect interface, i.e.\ 
for $r_k=1$, $r_v=1$, $Z=0$. 
For simplicity, in the case of a perfect interface we shall use  
the notation $k_{\rm{F}}=k_{\rm{F}}^{\rm{(N)}}=k_{\rm{F}}^{\rm{(S)}}$ 
and $E_{\rm{F}}=E_{\rm{F}}^{\rm{(N)}}=E_{\rm{F}}^{\rm{(S)}}$. 
From Eqs.\ (\ref{De}) and (\ref{S_NS}),  and in Andreev approximation 
($\Delta_0/E_{\rm{F}} \ll 1$, i.e.\  
$k^{(e)}_m \approx q^{(e)}_m$ except when they appear in an exponent), 
one finds 
\begin{equation}
\Phi_m(\varepsilon) = 2 k^{(e)}_m d. 
\label{box-Phi_m}
\end{equation} 
Thus, from Eq.\ (\ref{Phi_m}) the quantization condition becomes 
\begin{equation}
\left(k^{(e)}_m - k^{(h)}_m\right) d = n\pi 
+ \arccos (\varepsilon/\Delta_0).
\end{equation} 
Note that this result can also be derived from the 
Bohr-Sommerfeld quantization rule. In fact, $\hbar \Phi_m(\varepsilon)$ 
is the classical {\it action} for the electron moving along the $x$ 
direction between the superconductor and the wall at $x=d$. 

\begin{figure}[hbt]
{\centerline{\leavevmode \epsfxsize=7cm 
\epsffile{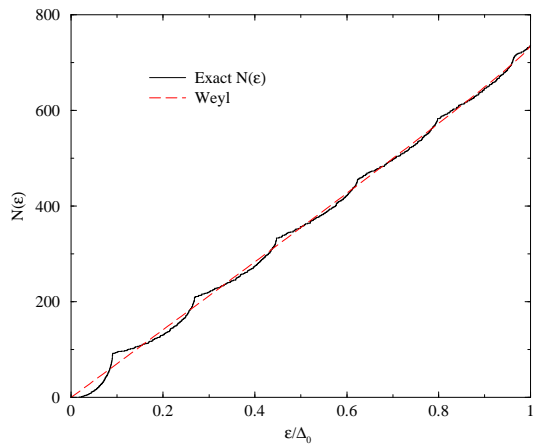}}}
\caption{
The counting function $N(\varepsilon)$ obtained from the exact quantum
mechanical calculations (solid line)
and the Weyl formula given in (\ref{Weyl}) (dashed line)
as functions of $\varepsilon/\Delta_0$ for the box geometry. 
The parameters are  $d/W = 3$, 
$k_{\rm{F}}W/\pi = 87.9$ and  
$\Delta_0/E_{\rm{F}}=0.02$.
\label{box-Weyl-A}}
\end{figure}

The exact counting function $N(\varepsilon)$ obtained (numerically) 
from Eq.\ (\ref{DNS}) and the Weyl formula given in Eq.\ (\ref{Weyl}) 
are plotted in Fig.\ \ref{box-Weyl-A}.  
For parameters see the figure caption.
One can see that the exact counting function oscillates around the 
curve obtained from our Weyl formula. 

\begin{figure}[hbt]
{\centerline{\leavevmode \epsfxsize=7cm 
\epsffile{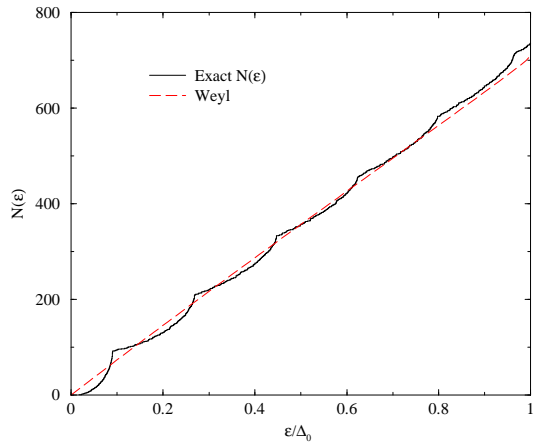}}}
\caption{
The exact counting function $N(\varepsilon)$ (solid line) 
and the analytical expression of the Weyl formula given 
in (\ref{box-Weyl-approx}) (dashed line) as functions of 
$\varepsilon/\Delta_0$ for the box geometry.
The parameters are the same as in Fig.\ \ref{box-Weyl-A}.
\label{box-Weyl-B}}
\end{figure}

To find an analytical expression for the Weyl formula, the sum 
in Eq.\ (\ref{Weyl}) is replaced by an integral and we get  
\begin{equation}
\tilde{N}(\varepsilon) = 
\frac{2}{\pi} \, \rho^{\rm{(N)}} \, E_{\rm{F}} \,  
g(\varepsilon/E_{\rm{F}}) + M_h
\left[1/2 - 1/\pi \, \arccos(\varepsilon/\Delta_0)\right],
\label{box-Weyl-approx}
\end{equation} 
where 
\begin{equation} 
g(x)= \sqrt{2x(1-x)} +(1+x) \arcsin \sqrt{(1-x)/(1+x)} 
-\pi/2\, (1-x),
\end{equation}
and  $\rho^{\rm{(N)}}=2m/\hbar^2\, A/(4\pi)$ is the density of states  
for the isolated N region of area $A=Wd$. 
The number of open channels for the hole-like
quasiparticles is $M_h = [k_{\rm{F}}W/\pi \, 
\sqrt{1-\varepsilon/E_{\rm{F}}}]$, where 
$[\cdots ]$ stands for the integer part.
For $x\ll 1 $ (typically $x<0.1$) 
$g(x)\approx \pi x$. Thus, the leading term of the smooth part of the 
counting function of Andreev states is 
$\tilde{N}(\varepsilon) \approx  2 \,
\varepsilon \, \rho^{\rm{(N)}}$. Electron-like and hole-like
quasiparticles make equal contributions to the Weyl formula 
(\ref{box-Weyl-approx}), thereby the factor 2.
The exact counting function $N(\varepsilon)$ and the analytical form of 
the Weyl formula given in Eq.\ (\ref{box-Weyl-approx}) are plotted 
in Fig.\ \ref{box-Weyl-B}.
It is seen from the figure that the analytical expression (dashed curve) 
deviates only for energies close to the gap. 

\subsection{Weyl formula for disk (no mismatch)}
\label{Weyl-disk}

In this section an analytical expression of the Weyl formula 
is derived for disk geometry.  
However, most of the results obtained in this section will be used in
Sec. \ref{sing-disk}, too.  
Again, we shall take the case of perfect 
NS interface, namely no mismatch and tunnel barrier are assumed 
($r_k=1$, $r_v=1$, $Z=0$).  
To find an analytical expression for the eigenphase
$\Phi_m(\varepsilon)$ 
defined in Eq.\ (\ref{S_NS}) we have to approximate the determinant 
$D^{\rm{(e)}}_m(\varepsilon)$ in (\ref{De-disk}).
The details of the approximations are presented in Appendix \ref{app-disk}.

\begin{figure}[hbt]
{\centerline{\leavevmode \epsfxsize=7cm 
\epsffile{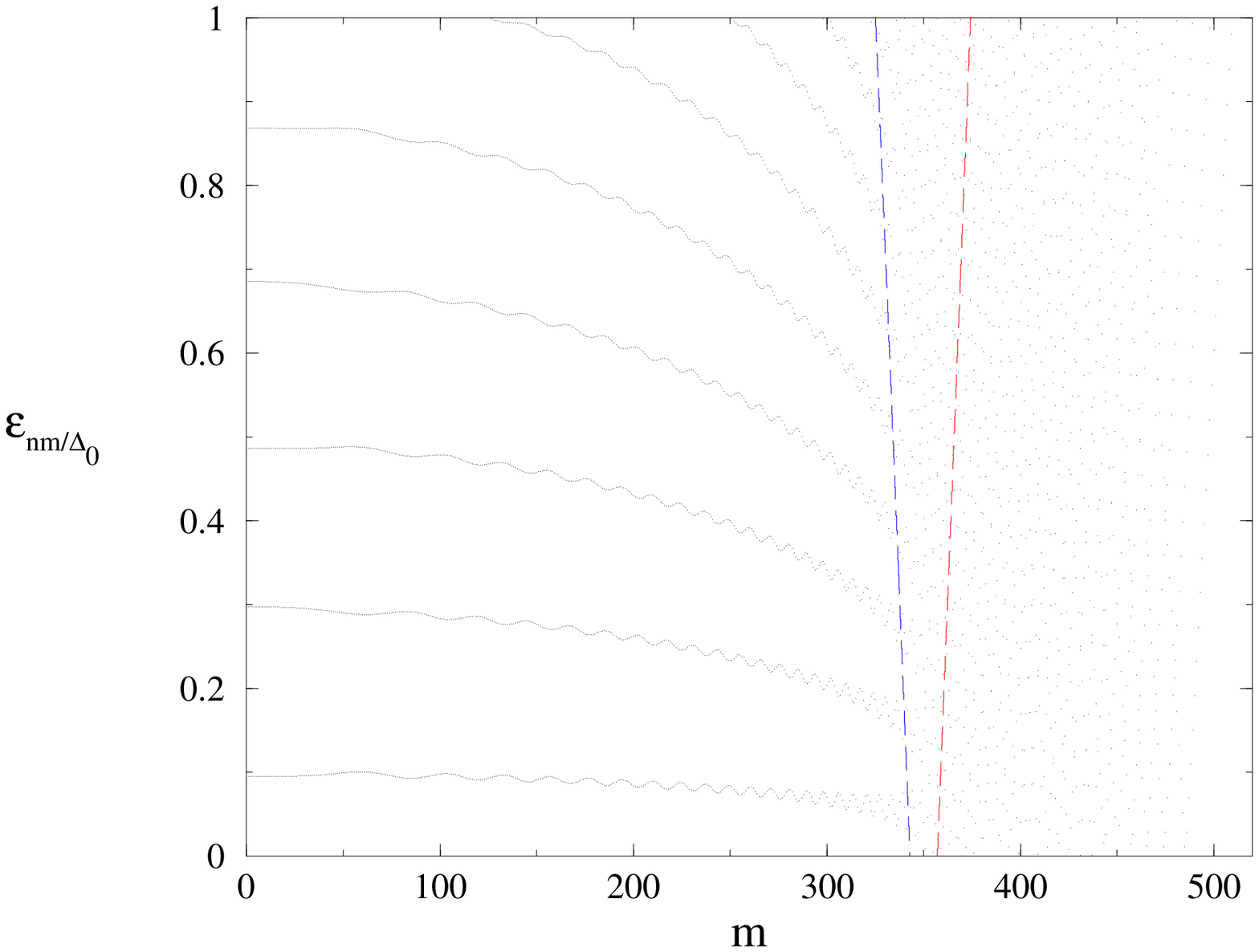}}}
\caption{The exact energy levels, $\varepsilon_{nm}$  (dots) in units of
$\Delta_0$ as a function of the angular momentum quantum number $m$. 
There is no mismatch, and the potential of the tunnel barrier is zero. 
The parameters are $R_{\rm{S}}/R_{\rm{N}}=0.7$, 
$\Delta_0/E_{\rm{F}}=0.1$, and 
$k_{\rm{F}}R_{\rm{S}}=350$.
The intermediate $m$ states are located between the two dashed lines.
The Andreev states are to the left of the left dashed line,  
and the whispering gallery modes are to the right of the right one.
\label{disk-raw}}
\end{figure}

\begin{figure}[hbt]
{\centerline{\leavevmode \epsfxsize=7cm 
\epsffile{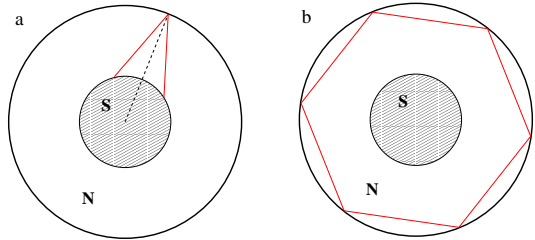}}}
\caption{The corresponding semiclassical orbits for (a) Andreev states and
(b) whispering gallery states.
\label{disk-AD-suttogo}}
\end{figure}

To summarize, according to the approximations of the determinants 
appearing in the secular equation (\ref{DNS}), three ranges of $m\ge 0$ 
can be distinguished: 
for Andreev states $m < k_h R_{\rm S}-\sqrt[3]{k_h R_{\rm S}}$, 
for whispering gallery states $m > k_e R_{\rm S}+\sqrt[3]{k_e R_{\rm S}}$, 
and for  intermediate states 
$k_h R_{\rm S}-\sqrt[3]{k_h R_{\rm S}} < 
m < k_e R_{\rm S}+\sqrt[3]{k_e R_{\rm S}}$. Owing to the degeneracy of the
$\pm m$ states the three ranges for $m\ge 0$ and $m \le 0 $ are
symmetrically located with respect to the state $m=0$.
This structure of the energy levels will be called a `mixed phase' (MP).   
The exact energy levels, $\varepsilon_{nm}$ 
obtained (numerically) from (\ref{DNS}) are shown as functions of 
the angular quantum number $m$ in Fig.\ \ref{disk-raw}. 
For each $m$ the solutions are labeled by the quantum
number $n$ (radial quantum number).  Since the pairs $\pm m$ are
degenerate,  only $m\ge0$ states are plotted in the figure. 
The two dashed lines in Fig.\ \ref{disk-raw} separate the three types of
states characterized by $m$. One can see that the intermediate states
indeed occupy a narrow range in $m$ compared to Andreev and
whispering gallery states. It is also clear from the figure that 
the energy levels for Andreev states `oscillate' with increasing 
amplitude as $m$ increases towards the onset of the intermediate states 
(left dashed line in the figure). 
Inserting  Eq.\ (\ref{Fi-kul}) into the quantization condition
given by  Eq.\ (\ref{Phi_m}) one can approximately calculate the energy 
levels for Andreev states. Plotting these energy levels in
Fig.\ \ref{disk-raw} (not shown in the figure) we found that the 
exact energy levels `oscillate' around these approximate ones.  
Similar behavior of the energy levels has been found by
\v{S}ipr et al.\ \cite{Gyorffy}.  
For whispering gallery states the approximate secular equation is given by 
Eq.\ (\ref{whisp-sec}), and its solutions coincide very accurately with
exact energy levels.
These approximate energy levels for Andreev and whispering gallery 
states are those which can be obtained from the semiclassical 
quantization of the Schr{\"o}dinger equation describing the radial motion of 
the electron\cite{Brack}.
Indeed, one can check that the first two terms for $\Phi_m(\varepsilon)$ 
in Eq.\ (\ref{phi-disk-1}) multiplied by $\hbar$ give 
the {\it radial action} for the electron moving between the two circles. 
The last three terms are constant, i.e.\ independent of  
$\varepsilon$ therefore they do not play any role in the dynamics of 
electron.
These trajectories of the electrons correspond to Andreev states.  
The semiclassical orbits for Andreev and whispering gallery states 
are shown in Fig.\ \ref{disk-AD-suttogo}.

Inserting  (\ref{Fi-kul}) into the Weyl formula (\ref{Weyl}) one can
determine  the contributions of the Andreev states 
to the smooth part of the counting function:
\begin{equation}
\tilde{N}_{\rm AS}(\varepsilon) = \frac{1}{\pi}\, 
\sum_{|m|<M_{\rm AS}} \, 
\left[\vartheta_m(k_e R_{\rm N}) -\vartheta_m(k_e R_{\rm S}) 
-\vartheta_m(k_h R_{\rm N}) +\vartheta_m(k_h R_{\rm S}) \right]
+2\, M_{\rm AS}\left(\frac{1}{2}-\frac{1}{\pi} \, 
\arccos \frac{\varepsilon}{\Delta_0} \right),
\label{N_AS}
\end{equation}
where $M_{\rm AS} = \left[k_h R_{\rm S}
-\sqrt[3]{k_h R_{\rm S}}\,\right]$ is the highest $m$ for the Andreev states. 

For whispering gallery modes the eigenphase $\Phi_m (\varepsilon)$ 
in (\ref{S_NS})  cannot be determined because
$D^{\rm{(e)}}_m(\varepsilon)$ is approximately zero. In this case 
the contribution to the smooth part of the 
counting function will be calculated from the secular equation 
(\ref{whisp-sec}) for whispering gallery states. 
It is well known that the zeros of the Bessel functions can be determined
to a good approximation by applying Debye's asymptotic expansion given in
Eq.\ (\ref{Bessel-Debye}). The solution of
the secular equation (\ref{whisp-sec}) 
for whispering gallery states is then equivalent to the solution of the
following pair of equations for $\varepsilon$: 
\begin{eqnarray}
F_m^{(e)}(\varepsilon) = \frac{1}{\pi} \, 
\vartheta_m(k_e R_{\rm N})-\frac{3}{4} & = & n_1 \\[1ex]  
F_m^{(h)}(\varepsilon) = \frac{1}{\pi} \, 
\vartheta_m(k_h R_{\rm N})-\frac{3}{4} & = & n_2,  
\end{eqnarray}
where $n_1, n_2 = 0,\pm 1, \pm 2, \cdots $, and $\vartheta_m(x)$ is given
by Eq.\ (\ref{eta-def}). From these equations one can find the density of
states of whispering gallery states much in the same way  
as from Eq.\ (\ref{rho-1}) for the Andreev states: 
\begin{equation}
\rho_{\rm{wgs}}(\varepsilon) = 
2 \sum_{m=M_{\rm wgs}}^{k_e R_{\rm N}}
\, \left | \frac{d F_m^{(e)}(\varepsilon)}{d \varepsilon}\right| +  
2 \sum_{m=M_{\rm wgs}}^{k_h R_{\rm N}}
\, \left | \frac{d F_m^{(h)}(\varepsilon)}{d \varepsilon}\right|,
\label{rho_wgs}
\end{equation}
where $M_{\rm wgs}= \left[ k_e R_{\rm S}+\sqrt[3]{k_e R_{\rm S}} \right ]$ 
is the smallest angular momentum quantum number $m$ for whispering 
gallery states, and the factor 2 corresponds to the $\pm m$ degeneracy.
The contribution of whispering gallery states to the 
smooth part of the counting function can be obtained from 
$\tilde{N}_{\rm{wgs}}(\varepsilon) = 
\int_0^\varepsilon \, 
\rho_{\rm {wgs}}(\varepsilon^\prime) \, d \varepsilon^\prime$ which yields 
\begin{equation}
\tilde{N}_{\rm{wgs}}(\varepsilon) = 
\frac{2}{\pi} \sum_{m=M_{\rm wgs}}^{k_e R_{\rm N}}
\vartheta_m(k_e R_{\rm N}) - 
\frac{2}{\pi} \sum_{m=M_{\rm wgs}}^{k_h R_{\rm N}}
\vartheta_m(k_h R_{\rm N}).
\label{N_wgs}
\end{equation}
The negative sign in front of the second term comes from the fact that 
the derivative of $\vartheta_m(k_h R_{\rm N})$ with respect to
$\varepsilon$ is negative. 

The counting function $\tilde{N}(\varepsilon)$ for the NS disk system 
is the sum of the contribution of Andreev states 
($\tilde{N}_{\rm AS}(\varepsilon)$) and that of whispering gallery states 
($\tilde{N}_{\rm{wgs}}(\varepsilon)$).
We now neglect the contribution of the intermediate states.
To find an analytical expression for $\tilde{N}(\varepsilon)$, 
the summation over $m$ is replaced by an integral 
in Eqs.\ (\ref{N_AS}) and (\ref{N_wgs}) 
and  for $k_{\rm F} R_{\rm S} \gg 1 $ 
$M_{\rm{AS}} \approx M_{\rm{wgs}} \approx k_h R_{\rm S}$ is taken.
The last approximation corresponds to replacing the intermediate
states by whispering gallery states. This is quite a good approximation,  
since the number of intermediate states is much smaller than that of 
the Andreev and whispering gallery states.   
The resulting integrals can be performed analytically. 
After some algebra we obtain 
\begin{equation}
\tilde{N}(\varepsilon) = 
\frac{k_{\rm F}^2 R_{\rm N}^2 }{2} \, 
\frac{\varepsilon}{E_{\rm{F}}} - 
\frac{k_{\rm F}^2 R_{\rm S}^2 }{2} \, 
g(\frac{\varepsilon}{E_{\rm{F}}}) 
+2 k_{\rm{F}}R_{\rm{S}} \, \sqrt{1- \frac{\varepsilon}{E_{\rm{F}}}}
\left (
\frac{1}{2} - \frac{1}{\pi} \arccos \frac{\varepsilon}{\Delta_0}
\right),
\label{NS-disk-Weyl}
\end{equation}
where 
\begin{equation}
g(x) = \frac{3}{\pi} \sqrt{2x(1-x)} 
+ \frac{1}{\pi}(1+x) \arcsin \sqrt{\frac{1-x}{1+x}} - 
(1-x) \left( \frac{1}{2} +\frac{2}{\pi}\arccos\sqrt{\frac{1-x}{1+x}}\right).
\end{equation}
This is our Weyl formula for NS concentric disk systems. 
Usually, ${\varepsilon}/{E_{\rm{F}}} \ll 1$ and for $x \ll 1$,  
$g(x) = x$ in leading order, therefore the first two terms 
in Eq.\ (\ref{NS-disk-Weyl}) yield $2 \rho^{\rm{(N)}} \varepsilon$, 
where $\rho^{\rm{(N)}}=2m/\hbar^2\, A/(4\pi)$ is 
the density of states for a normal annular billiard of area 
$A=\left(R_{\rm N}^2 - R_{\rm S}^2\right)\pi$.
The factor 2 comes from the electron/hole contributions for NS systems.
The higher order corrections in $g(x)$ become more relevant with increasing
$\varepsilon$.  
The last term in (\ref{NS-disk-Weyl}) is related to the phase shift due to
the Andreev reflection at the NS interface. It is important to stress that 
the Weyl formula given in Eq.\ (\ref{NS-disk-Weyl}) was
derived for the case $R_{\rm S} \gg \xi (\varepsilon)$, i.e.\  when 
Andereev states and whispering gallery states coexist (mixed phase). 
In the opposite case, as we shall see in Sec.\ \ref{fazis-diag}, 
no Andreev states exist. 

\begin{figure}[hbt]
{\centerline{\leavevmode \epsfxsize=7cm 
\epsffile{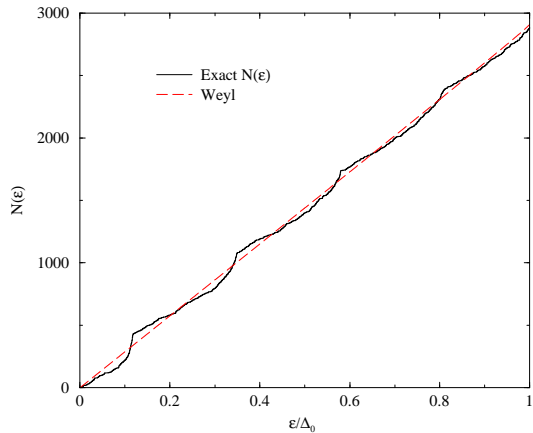}}}
\caption{The exact counting function $N(\varepsilon)$ obtained from 
Eqs.\ (\ref{DNS}) (solid line) 
and $\tilde{N}(\varepsilon)$ given by (\ref{NS-disk-Weyl}) (dashed line) 
as functions of $\varepsilon/\Delta_0$ for the disk geometry. 
There is no mismatch and the potential of the tunnel barrier is zero. 
The parameters are: $R_{\rm{S}}/R_{\rm{N}}=2/7$, 
$\Delta_0/E_{\rm{F}}=0.05$ and 
$k_{\rm{F}}R_{\rm{S}}=100$. 
\label{disk-Weyl-1}}
\end{figure}

In Fig.\ \ref{disk-Weyl-1} the exact counting function and 
the Weyl formula given by Eq.\ (\ref{NS-disk-Weyl}) are plotted 
as functions of $\varepsilon/\Delta_0$. For parameters see the figure
caption.
The singularities of the DOS (derivative of the counting function with
respect to $\varepsilon$) will be discussed in the next section.

\section{Singularities in DOS}
\label{sing}

The exact counting function, $N(\varepsilon)$ has {\it cusps\/} at
some energies as shown 
Figs.\ \ref{box-Weyl-A} and \ref{disk-Weyl-1} 
for box and disk geometries, and perfect interface.
This implies that the DOS of the Andreev states is {\it discontinuous\/}
here. Moreover, these cusps are at {\it equal distances\/}.
Similar singularities of the DOS for NS systems have already been 
found by de Gennes et al.\ \cite{deGennes-Saint-James} as well as 
in the numerical works of Richter et al.\ \cite{Richter1} 
(for NS systems), Nazarov et al.\ \cite{Nazarov} (for Andreev
billiard), and \v{S}ipr et al.\ \cite{Gyorffy} (for SNS systems).
In this section we investigate the singularities in the DOS for the
NS systems shown in Fig.\ \ref{abra-gen-box-disk}. 

Several authors \cite{Melsen,Nazarov,Heny,Richter1} have 
already derived the Bohr-Sommerfeld approximation to the density of states,
\begin{equation}
\rho_{\rm BS}(\varepsilon )= M \int_0^\infty ds \, P(s) \sum_{n=0}^\infty \, 
\delta \left(\varepsilon - 
\left( n+\frac{1}{2}\right)\frac{\pi \hbar v_{\rm F}}{s} \right),
\label{BS-rho}
\end{equation}
where $M$ is the number of modes in the superconducting leads 
connected to the normal billiard, and $P(s)$ is the classical probability 
that an electron entering the billiard exits after a path length $s$. 
Starting from our quantization rule given in Eq.\ (\ref{Phi_m}) 
we re-derive Eq.\ (\ref{BS-rho})  in this section, and give an explicit 
expression of the probability $P(s)$ in terms of $\Phi_m(\varepsilon)$.
We shall show that $P(s)$ is singular at some $s$ resulting in the 
singularities of the DOS. These singularities correspond to 
the cusps in the integrated DOS $N(\varepsilon)$. 
It turns out that the simple Bohr-Sommerfeld approximation to the DOS 
is a good approach to understand the singularities in the DOS provided 
that $P(s)$ is approximated correctly for small $s$.   

The solution of Eq.\ (\ref{Phi_m}) gives the discrete energy levels for
the NS systems. 
Since $\varepsilon \ll E_{\rm F}$, one can Taylor expand the LHS of 
Eq.\ (\ref{Phi_m}) in terms of $\varepsilon$ and find 
\begin{equation}    
\varepsilon_{n,m} = 
\frac{\left(n+\frac{1}{2}\right)\pi}{\Phi_m^\prime(0)},
\label{E_mn} 
\end{equation}
where $\Phi_m^\prime(0)$ denotes the derivative of $\Phi_m(\varepsilon)$ 
with respect to $\varepsilon$, evaluated 
at the Fermi energy, i.e.\ $\varepsilon =0$. For the NS box systems 
$m=1,2,\dots,M$, where the energy dependent $M$ is 
the number of `transverse modes'. 
For the NS disk systems $|m|<M$, where $M$ is the maximum of the angular
momentum quantum number $m$ for the Andreev states.   
We consider only the positive energy spectrum, therefore $n\ge 0$.
Here we take the box geometry. The following expressions for disk geometry
can be obtained straightforwardly by taking into account the fact that the
set of $m$ is different in this case. Thus $M$ has a different meaning
for box and disk geometries. 
We shall always indicate how to convert the box geometry results to the
disk geometry case. 
To get Eq.\ (\ref{E_mn}), the phase term in Eq.\ (\ref{Phi_m}) 
due to Andreev reflection, $\arccos (\varepsilon/\Delta_0)$ was 
approximated by $\pi/2$, which is valid for $\varepsilon \rightarrow 0$. 
Later a better approximation will be given by Taylor expanding 
the phase term, too. The density of states is then 
\begin{equation}
\rho (\varepsilon ) = 
\sum_{n=0}^\infty\sum_{m=1}^M \, \delta (\varepsilon -\varepsilon_{n,m} ) =
\sum_{n=0}^\infty\sum_{m=1}^M \, \int \, ds \, \delta \left(\varepsilon - 
\frac{\left(n+\frac{1}{2}\right)\hbar \pi v_{\rm F}}{s}\right) 
\delta \left(s-\hbar v_{\rm F} \Phi_m^\prime(0)\right).
\label{rho-BS-1}
\end{equation}
We have seen in Secs.\ \ref{box-Weyl} and \ref{Weyl-disk} that 
$\hbar\Phi_m(0)$ is the classical action for the electron 
with Fermi energy moving in the N region between two subsequent bounces
at the superconductor. Thus, $s=\hbar v_{\rm F} \Phi_m^\prime(0)$ is 
the path length of the trajectories of the electron between two 
successive bounces at the NS interface. 
We now show that the above expression for $\rho (\varepsilon )$ 
can be rewritten in the same form as in (\ref{BS-rho}), which makes 
it possible to express the probability $P(s)$ 
in terms of $\Phi_m(\varepsilon)$.

Applying the Poisson formula\cite{Berry-1,Brack} for the summation over 
$m$ in (\ref{rho-BS-1})
we have
\begin{equation}
\rho (\varepsilon) = 
\sum_{n=0}^\infty \, \int_0^\infty ds \, \delta \left(\varepsilon - 
\frac{\left(n+\frac{1}{2}\right)\hbar \pi v_{\rm F}}{s}\right) 
\sum_{k=-\infty}^{\infty} \, \int_{1/2}^{M+1/2} dm \, 
\delta \left(s-\hbar v_{\rm F} \Phi_m^\prime(0)\right) e^{i2\pi km}.
\end{equation}
Keeping only the non-oscillating term ($k=0$) in the sum over $k$, and then
performing the integral over $m$ we obtain the same Bohr-Sommerfeld
approximation to the DOS as in Eq.\ (\ref{BS-rho}). The probability
$P(s)$ is given by 
\begin{equation}
P(s) = \frac{\Theta\left(M-m^*\right)\Theta\left(m^* -1\right)}
{M \hbar  v_{\rm F} \,  
\left| \frac{\partial\Phi_m^\prime(0)}{\partial m}\rule[-1.6ex]{.2pt}{4ex}\;
        \raisebox{-1.5ex}{$\scriptstyle {m=m^*}$}\right|},
\label{Ps}
\end{equation}
where the $s$-dependent $m^*$ satisfies
\begin{equation}
\hbar  v_{\rm F} \Phi_m^\prime(0)\rule[-1.6ex]{.2pt}{4ex}\;
        \raisebox{-1.5ex}{$\scriptstyle {m=m^*}$} = s.
\label{mcsillag}
\end{equation}
$\Theta(x)$ is the Heaviside function. 
Note that the probability $P(s)$ is normalized to 1, i.e.\ 
$\int_0^\infty P(s)\, ds =1$. For disk geometry the numerator in
(\ref{Ps}) has to be replaced by $\Theta(M-|m^*|)/2$, and the necessary 
replacement in (\ref{BS-rho}) is $M\rightarrow 2M$.
Our main result is that the probability $P(s)$ is expressed in terms of 
$\Phi_m(\varepsilon)$, which has already been determined for box and disk
geometries of the NS system 
(see Eqs.\ (\ref{box-Phi_m}) and (\ref{phi-disk-1})). 

Performing the integral over $s$ in Eq.\ (\ref{BS-rho}) one finds
\begin{equation}
\rho_{\rm BS}(\varepsilon )= \frac{M}{\varepsilon} \, 
\sum_{n=0}^\infty \,  s_n(\varepsilon) P(s_n(\varepsilon)),
\label{BS-rho-b}
\end{equation} 
where 
\begin{equation}
s_n(\varepsilon) = 
\frac{\left(n+\frac{1}{2}\right)\pi \hbar v_{\rm F}}{\varepsilon}
=\frac{\left(n+\frac{1}{2}\right)2\pi} {k_{\rm F}} \, 
\frac{E_{\rm F}}{\varepsilon}.
\label{sn-eps}
\end{equation}
Similarly, the counting function $N_{\rm BS}(\varepsilon) = 
\int_0^\varepsilon d \varepsilon^\prime \, 
\rho_{\rm BS}(\varepsilon^\prime)$ can be easily found
\begin{equation}
N_{\rm BS}(\varepsilon) = 
M \sum_{n=0}^\infty \, \int_{s_n(\varepsilon)}^\infty \, 
P(s)\, ds.  
\label{BS-Ne}
\end{equation}
Both in (\ref{BS-rho-b}) and (\ref{BS-Ne}) we have to replace $M$ by $2M$
in the case of NS disk systems as $M$ has a different meaning in this case.
The above results are valid only for small $\varepsilon$. 
Numerical results show that a better approximation of the counting function
can be obtained by Taylor expanding the phase shift related to the Andreev 
reflection: 
$\arccos (\varepsilon / \Delta_0) \approx \pi/2 - \varepsilon / \Delta_0$.   
Then the discrete energy levels are given by 
\begin{equation}
\varepsilon_{n,m} = 
\frac{\left(n+\frac{1}{2}\right)\pi}{\Phi_m^\prime(0)+\frac{1}{\Delta_0}}. 
\end{equation}
Carrying out the same procedures as before one finds that the 
counting function $N_{\rm BS}(\varepsilon)$ is still given 
by Eq.\ (\ref{BS-Ne}) after the replacement 
\begin{equation}
s_n(\varepsilon) \rightarrow s_n(\varepsilon) -
\xi_0
\label{sn-eps-rep}
\end{equation}
is made in (\ref{sn-eps}). 
Here $\xi_0 = \xi(\varepsilon=0)=\hbar v_{\rm F}/\Delta_0
=2E_{\rm F}/(k_{\rm F}\Delta_0)$ is 
the coherence length at the Fermi energy (see Eq.\ (\ref{coherence-xi})).

In the following two subsections we shall calculate 
$P(s)$ and the counting function $N_{\rm BS}(\varepsilon)$ 
for NS box and disk systems. We shall see that the probability $P(s)$, 
more precisely $sP(s)$, 
is singular at some $s_{\rm sing}$ for both NS systems. 
Then, if $s_{\rm sing}$ is known, one can determine 
from (\ref{sn-eps-rep}) the energies where $P(s_n(\varepsilon))$---and  
consequently (see Eq.\ (\ref{BS-rho-b})) the density of states 
$\rho_{\rm BS}(\varepsilon )$---is singular:
\begin{equation}
\varepsilon_n^{\left({\rm sing}\right)} = 
\frac{\left(n+ {1}/{2}\right)\pi}
{1+ s_{\rm sing}/\xi_0}\, \Delta_0,
\label{ep-sing}
\end{equation}
which is valid for all $n$ for which 
$\varepsilon_n^{\left({\rm sing}\right)} < \Delta_0$. 
Note that this expression can be applied to those normal systems attached
to a superconductor for which $sP(s)$ is singular. 
It is clear from this expression that these singularities are 
at {\it equal} distances. The analytical behavior
of $P(s)$, and thus, the existence of the singularities in the DOS 
for perfect NS interface inherently depends on the geometry of 
the isolated normal billiard. Once $sP(s)$ is a singular function of $s$ 
(because of the shape of the normal region), then the subsequent
singularities in the DOS are at equal distances.   
We shall see that from the exact calculation of the energy eigenvalues 
for NS box and disk systems the positions of the singularities in the DOS
agree very well with those given by the above derived approximate  
expression. We would like to emphasis that the second term in 
the denominator of  Eq.\ (\ref{ep-sing}) resulted from the Taylor 
expansion of $\arccos(\varepsilon/\Delta_0)$ which is 
the phase shift due to the Andreev reflection. We shall see below 
that this extra term involving the coherence length $\xi_0$ 
in Eq.\ (\ref{ep-sing}) is necessary for getting good agreements  
between the exact and that of predicted by Eq.\ (\ref{ep-sing}) 
for the position of the singularities.

\subsection{Singularities in DOS for Box}
\label{sing-box}

\begin{figure}[hbt]
{\centerline{\leavevmode \epsfxsize=7cm \epsffile{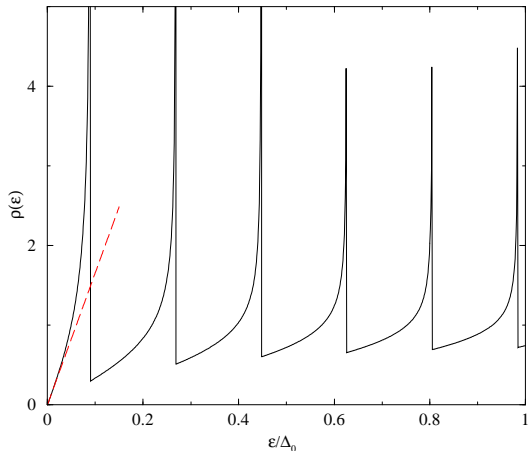}}}
\caption{The Bohr-Sommerfeld approximation of the DOS for the box
geometry---normalized by the
DOS of a normal billiard of area $A=dW$ (solid line)---and its slope 
for $\varepsilon \rightarrow 0$ (dashed line) 
as functions of $\varepsilon/\Delta_0$.
The parameters are the same as in Fig.\ \ref{box-Weyl-A}. With these
parameters $2d/\xi_0 = 16.57$.
\label{abra-box-rho}}
\end{figure}

In Sec.\ \ref{box-Weyl} we have calculated $\Phi_m(\varepsilon)$ 
for the NS box system with perfect interface.
Inserting  $\Phi_m(\varepsilon)$ given by (\ref{box-Phi_m}) into 
Eqs.\ (\ref{Ps}) and (\ref{mcsillag}) one finds
\begin{equation}
P(s) = \frac{4 d^2}{s^3\sqrt{1-{\left(\frac{2d}{s}\right)}^2}}\, 
\Theta (s- s_{\rm min}), 
\label{Ps-box}
\end{equation}
where $s_{\rm min}=2d/\sqrt{1-1/M^2}$ and 
$M=k_{\rm F}W/\pi$ is the number of `transverse modes'.
For a large number of transverse modes, i.e.\ $M\gg 1$ 
we have $s_{\rm min}=2d$. 
Notice that the same result can be found for $P(s)$ by simple geometrical
considerations\cite{Kaufmann}.
$P(s)$ is singular at $s_{\rm sing}=2d$, therefore the DOS
is singular at the energies given by (\ref{ep-sing}). The DOS can be
calculated from Eq.\ (\ref{BS-rho-b}) in Bohr-Sommerfeld approximation.  
In Fig.\ \ref{abra-box-rho} the normalized DOS, 
$\rho_{\rm BS}(\varepsilon)/(2\rho^{({\rm N})})$ is plotted 
in Bohr-Sommerfeld approximation together with its slope 
for $\varepsilon \rightarrow 0$. 
Here $\rho^{({\rm N})}=2m/\hbar^2(A/4\pi)$ is the DOS
of the normal box of area $A=dW$.   
From (\ref{sn-eps}) it is clear that the DOS  for small $\varepsilon$ 
is dominated by the large $s$ behavior of $P(s)$. 
According to (\ref{Ps-box}), $P(s) \rightarrow 4d^2 s^{-3}$ in this case. 
Thus, 
\begin{equation}
\frac{\rho_{\rm{BS}} (\varepsilon)}{ 2\rho^{\rm{(N)}}} \rightarrow  
\frac{\varepsilon}{\pi E_T}, \,\,\,\, 
{\rm for}\,\,\,\,\varepsilon\rightarrow 0,
\end{equation}
where 
$\rho^{\rm{(N)}}= \left(2m_{\rm{N}}/\hbar^2\right)A/(4\pi)$
is the DOS for the isolated billiard of area $A=dW$ and
$E_T=M/(4\pi \rho^{\rm{(N)}})$ is the Thouless energy defined 
in Refs.~\onlinecite{Melsen,Richter1}.
This result is shown by the dashed line in Fig.\ \ref{abra-box-rho}.
For small energies $\varepsilon$ the DOS is proportional 
to $\varepsilon$ in agreement with the findings of 
Melsen et al.\ \cite{Melsen,Melsen-2} and Ihra et al.\ \cite{Richter1}. 
However, the slope of ${\rho_{\rm{BS}} (\varepsilon)}/{ 2\rho^{\rm{(N)}}}$
is less than what these authors found. The reason is that the
geometries of the billiards they studied are slightly different from our box
geometry. Namely, the width $w$ of their superconducting lead is 
smaller than the side length of the billiard with which it is in contact.
In these geometries, by quantum mechanical calculations, 
Ihra et al.\ \cite{Richter1} found 3 pronounced peaks 
in the DOS approximately at $\varepsilon/E_{\rm T}= 3.1, 8.3, 13.4$  
(see Fig.\ 3. in that paper). 
They mention that similar peaks were observed for other parameter
values of the billiards. 
If we assume that these peaks are indeed singularities in the DOS then
fitting these data to Eq.\ (\ref{ep-sing}) one finds that 
$P(s)$ is singular at $s=s_{\rm sing} \approx 1.1 L_{\rm T}$, 
where $L_{\rm T}=\pi a^2/w$ is the Thouless length used in their paper. 
The Thouless length is related to the mean escape time by 
$\tau_{\rm esc}=L_{\rm T}/v_{\rm F}$ for the billiard which is open along
the superconducting lead. Since it is reasonable to expect  
$P(s)$ to be high around $s=L_{\rm T}$, our theory about the
singularities in the DOS may explain the reason for the pronounced peaks 
in the DOS observed by Ihra and his coworkers\cite{Richter1}. 
They have calculated $P(s)$ for another set of parameters of the billiard, 
and a peak can also be seen around $s=L_{\rm T}$ in Fig.\ 2.\ in their work. 
However, to confirm the existence of the singularity in $P(s)$
one needs to calculate $P(s)$ on a finer scale than in Fig.\ 2. of their
paper. 

\begin{figure}[hbt]
{\centerline{\leavevmode \epsfxsize=7cm \epsffile{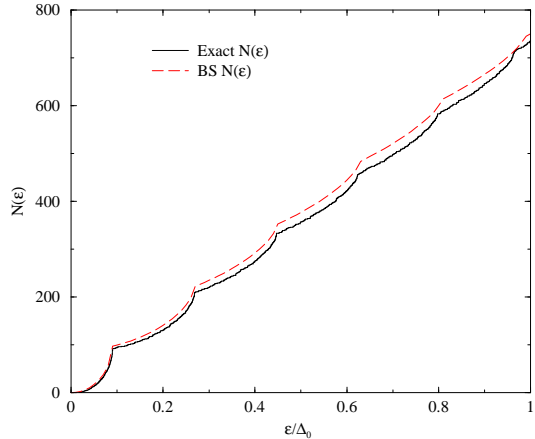}}}
\caption{The exact counting function 
$N(\varepsilon)$ (solid line) and 
$N_{\rm BS}(\varepsilon)$ given in (\ref{box-BS}) (dashed line)  
as functions of $\varepsilon/\Delta_0$ for the box geometry.
The parameters are the same as in Fig.\ \ref{box-Weyl-A}.
\label{abra-Box-BS}}
\end{figure}

To avoid the errors of numerical differentiation, we shall compare the 
counting function obtained from the quantum mechanical calculations (see
Eq.\ (\ref{DNS})) with the counting function 
$N_{\rm BS}(\varepsilon)$ in Bohr-Sommerfeld approximation. 
Using (\ref{BS-Ne}) and (\ref{Ps-box}),  
and performing the integration, we obtain 
\begin{equation}
N_{\rm BS}(\varepsilon) = M \sum_{n=0}^\infty f(s_n(\varepsilon)),
\label{box-BS}
\end{equation}
where
\begin{equation}
 f(s) = \left\{ \begin{array}{ll}
 1,  & \mbox{if}\,\,\,\, s \le 2d,  \\
 1-\sqrt{1-4d^2/s^2},& \mbox{if} \,\,\,\, s > 2d,
\end{array}   
 \right.  
\end{equation}
and $s_n(\varepsilon)$ is given in Eq.\ (\ref{sn-eps-rep}).
The above derived counting function $N_{\rm BS}(\varepsilon)$ is 
plotted in Fig.\ \ref{abra-Box-BS}  together with the exact
counting function. 
One can see that the agreement is excellent at low energies (below the
second cusp) but for higher
energies the exact counting function is smaller than the approximated
one. However, the positions of the singularities agree quite well   
in the two curves. Slight deviations are observed only for the last 
two cusps. The same was found for other parameter ranges. 
Thus, our expression (\ref{ep-sing}) for the positions of the singularities 
in the DOS results in an excellent agreement with  exact quantum mechanical
calculations. 
Our result may highlight the origin of the singularities of the
DOS. One can see that $P(s)$ is singular at $s=2d$, which corresponds 
to the classical orbits when the electron enters the box parallel to 
the $x$ axis and is reflected back at the wall. The enhanced return
probabilities for these orbits are quite obvious. 
These are the orbits which result in the singularities of the DOS. 
It is also clear from this expression that the singularities are 
at equal distances. 

\begin{figure}[hbt]
{\centerline{\leavevmode \epsfxsize=7cm \epsffile{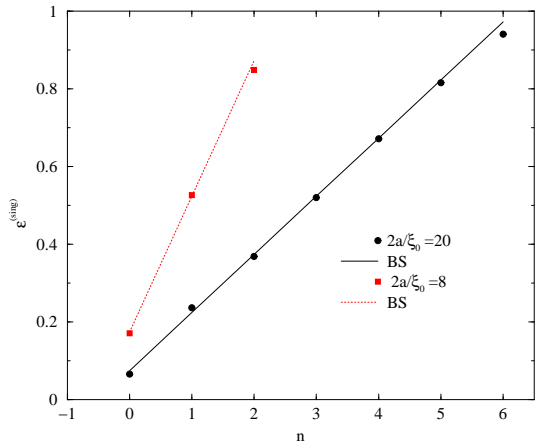}}}
\caption{The positions (in units of $\Delta_0$) of the singularities  
of the DOS obtained from Fig.\ 1 in the work of   
de~Gennes and Saint-James \cite{deGennes-Saint-James} (circles and squares) 
and from our expression (\ref{ep-sing}) (for clarity, the points are 
connected by lines). 
\label{abra-deGennes}}
\end{figure}

The expression for the positions of the singularities given 
by  Eq.\ (\ref{ep-sing}) can be applied to other NS systems, e.g.\  
that studied by de~Gennes and Saint-James \cite{deGennes-Saint-James}.
They considered an NS system in which a normal film of width $a$ is in
contact with a semi-infinite superconductor. 
This geometry of the NS system is similar to our NS box system shown  
in Fig.\ \ref{abra-gen-box-disk}a. 
The authors also found singularities in the DOS for 
parameter values $2a/\xi_0=2.5, 8, 20$. 
Using Fig.\ 1 of their paper we measured the positions of the
singularities in the DOS, labeled them as $n=0,1,2,\cdots$, and 
plotted them in Fig.\ \ref{abra-deGennes}
(circles and squares) against $n$. For  $2a/\xi_0=2.5$ they found
only one singularity which is not shown in Fig.\ \ref{abra-deGennes}. 
We can expect that the probability $P(s)$ for this system, similarly to
our box geometry, is singular at $s=2a$. Thus, assuming that 
$s_{\rm sing}=2a$  we calculated the positions of
the singularities from Eq.\ (\ref{ep-sing}) and also plotted them 
in Fig.\ \ref{abra-deGennes}.  For clarity, these points are connected by
lines.  
It can be seen from the figure that the agreement is excellent. Even for 
$2a/\xi_0=2.5$ (when there is only one singularity) the position 
of the singularity agrees very well with that found from 
our expression (\ref{ep-sing}). We note that 
to achieve such a good agreement the Taylor expansion of the phase shift 
$\arccos(\varepsilon/\Delta_0)$ in terms of $\varepsilon$ was necessary. 
Therefore, to understand the nature of the singularities one needs to take 
into account not only the singular behavior of P(s) which depends 
on the geometry of the billiard but a more physically related 
quantity, namely the coherence length of the superconductor.

Finally, we mention that Lodder and Nazarov\cite{Nazarov}  also found 
singularities in the DOS for different shapes of Andreev billiards. Our
results suggest that these singularities are related to some special 
classical orbits which are characteristic of the specific geometry.
\v{S}ipr et al.\ \cite{Gyorffy}  studied SNS systems taking 
fully into account the motion parallel to the infinite interface. 
Their results also show singularities in the DOS. We believe that 
these singularities are also related to some special classical orbits
which give rise to a singular behavior of the probability $P(s)$, similarly 
to the NS systems studied in this paper.    

\begin{figure}[hbt]
{\centerline{\leavevmode \epsfxsize=7cm \epsffile{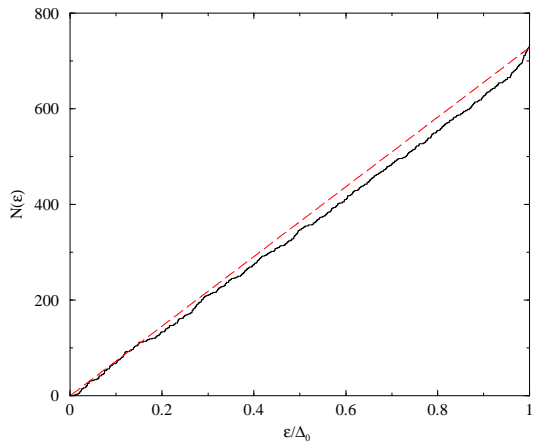}}}
\caption{The counting function $N(\varepsilon)$ as a
function of $\varepsilon/\Delta_0$ (solid line) 
for the box geometry when the interface is not perfect. 
The parameters are: $d/W = 3$, $k_{\rm{F}}^{\rm{(N)}}W/\pi = 87.9$,   
$\Delta_0/E_{\rm{F}}^{\rm{(N)}}=0.02$, $r_k=0.007$, $r_v=0.1$ and
with no tunnel barrier, i.e.\ $Z=0$. The Weyl formula in leading
order (see text) is given by the dashed line.
\label{abra-mismatch-box}}
\end{figure}

We also calculated the energy levels in the case of non-perfect NS
interface.  
Solving the secular equation (\ref{DNS}) numerically, the obtained 
counting function 
$N(\varepsilon)$ is shown in Fig.\ \ref{abra-mismatch-box}. 
We have used the same parameters for describing the mismatch 
in the Fermi energies and the effective masses as 
in Ref.~\onlinecite{Mortensen1}, namely 
$r_k=k_{\rm{F}}^{\rm{(N)}}/k_{\rm{F}}^{\rm{(S)}}$, 
$r_v=v_{\rm{F}}^{\rm{(N)}}/v_{\rm{F}}^{\rm{(S)}}$. 
The strength of the tunnel barrier is given by $Z$.
It can be seen from the figure that for a non-perfect NS interface 
the cusps in the counting function are 'washed out'.  It is related  
to the fact that the normal reflections at the interface are enhanced. 
The same is true for $Z\neq 0$.   
In Sec.\ \ref{box-Weyl} we have found that in leading order the Weyl
formula is given by $\tilde{N}(\varepsilon) \approx  2 \,
\varepsilon \, \rho^{\rm{(N)}}$, where 
$\rho^{\rm{(N)}}=2m/\hbar^2\, (A/4\pi)$ is the density of states  
for the isolated N region of area $A=Wd$. This is shown by the dashed line 
in Fig.\ \ref{abra-mismatch-box}. One can see from the figure that the 
exact counting function can be approximated by the above Weyl formula. This
implies that for non-perfect NS interfaces the effect due to the 
Andreev reflection is less significant, and the system behaves as a normal
metal regarding the energy levels.

\subsection{Singularities in DOS for disk}
\label{sing-disk}

The maximum angular momentum quantum number, $M_{\rm AS}$ for Andreev states
has been determined in Sec.\ \ref{Weyl-disk} after Eq.\ (\ref{N_AS}).
To have a better physical picture of these states we now give an 
estimation for $M_{\rm AS}$ based on semiclassical considerations. 
It is easy to see that the angular momentum (in units of $\hbar$) of 
the rays shown in  Fig.\ \ref{disk-AD-suttogo}a is 
$m=k_{\rm F}R_{\rm N}\sin (\alpha/2)$,
where $\alpha$ is the angle between the two rays at the outer circle. 
Rays that are tangent to the inner circle have maximal angular
momentum, in which case $\sin (\alpha/2)=R_{\rm S}/R_{\rm N}$. These are
the trajectories which still reach the superconductor so that they belong
to the Andreev states. 
Thus, the maximal angular momentum $m$ for the Andreev states 
is $k_{\rm F}R_{\rm S}$. This is a good estimate for  
$M_{\rm AS}$ given after Eq.\ (\ref{N_AS}).

In subsection \ref{Weyl-disk} we have seen that for 
$R_{\rm S} \gg \xi(\varepsilon)$ and for perfect NS interface 
the energy levels for NS disk systems can be classified into 
Andreev and whispering gallery states to a good approximation. 
We first consider the Andreev states. Inserting 
$\Phi_m(\varepsilon)$ given by Eq.\ (\ref{phi-disk-1}) into 
Eqs.\ (\ref{Ps}) and (\ref{mcsillag}), one finds after some algebra  
\begin{equation}
P(s) = \frac{1}{4\, s^2\, R_{\rm S}}
\frac{s_{\rm max}^4 -s^4}
{\sqrt{
\left[s_{\rm max}^4/s_{\rm min}^2 - s^2 \right]
\left[s^2 - s_{\rm min}^2 \right]
}}\,
\Theta(s_{\rm max}-s) \, \Theta(s-s_{\rm min}),  
\label{Ps-disk}
\end{equation}
where $s_{\rm min}=2(R_{\rm N}-R_{\rm S})$ and 
$s_{\rm max}=2\sqrt{R_{\rm N}^2-R_{\rm S}^2}$.
The two Heaviside functions come from the factor 
$\Theta(M-|m^*|)/2$ (see the text after Eq.\ (\ref{mcsillag})) when 
we change to the variable $s$.   
Here $s_{\rm min}$ and $s_{\rm max}$ correspond semiclassically to 
the smallest and the largest path lengths of the orbits related to 
the Andreev states. Obviously, $s_{\rm min}$ equals to the path length of
those orbits for which the trajectory of the electrons between two 
successive bounces at the NS interface is along the radius of the two
concentric circles. The trajectories of the electrons corresponding to the 
maximal path length for Andreev states are those which are tangent to the
inner circle of radius $R_{\rm S}$. 
The probability $P(s)$ is zero for $s_{\rm min}>s>s_{\rm max}$ and 
normalized to one. Note that in deriving
(\ref{Ps-disk}), the solution of Eq.\ (\ref{mcsillag}) is twofold for $m^*$ 
resulting in an extra factor 2 in $P(s)$.  
Notice that the same result can be found for $P(s)$ by simple geometrical
considerations\cite{Kaufmann}.
One can see that $P(s)$ is singular at $s=s_{\rm min}$.
The contributions of the Andreev states to the DOS in Bohr-Sommerfeld
approximation is given by Eq.\ (\ref{BS-rho-b}), which reads  
\begin{equation}
\rho_{\rm AD}(\varepsilon) = \frac{2k_{\rm F} R_{\rm S}}{\varepsilon}
\sum_{n=0}^\infty \,  s_n(\varepsilon) P(s_n(\varepsilon)),
\label{BS-rho-AD} 
\end{equation}
where $P(s)$ is given by (\ref{Ps-disk}) and $s_n(\varepsilon)$ by 
(\ref{sn-eps-rep}).
Since there is a maximum path length $s$ for Andreev states, 
the sum over $n$ in (\ref{BS-rho-b}) is indeed finite.

Next, we consider the contribution of the whispering gallery states 
to the DOS. 
After differentiations with respect to 
$\varepsilon$ in Eq.\ (\ref{rho_wgs}) and replacing 
the summations over $m$ by integrals the calculations 
can be carried out analytically (here we used 
$M_{\rm wgs}\approx k_e R_{\rm S}$  and neglected the
higher order terms in $\varepsilon/E_{\rm F}$). 
It is found that the DOS 
is {\it constant} in $\varepsilon$ and is given by 
\begin{equation}
\rho_{\rm wgs} = 
\frac{{\left(k_{\rm F}R_{\rm N}\right)}^2}{2E_{\rm F}}
\left[ 
1-\frac{2}{\pi}\left( 
\frac{R_{\rm S}}{R_{\rm N}} \, 
\sqrt{1-{\left(\frac{R_{\rm S}}{R_{\rm N}}\right)}^2} +
\arcsin \frac{R_{\rm S}}{R_{\rm N}} \right) \right].
\label{rho-wgs}
\end{equation}

\begin{figure}[hbt]
{\centerline{\leavevmode \epsfxsize=7cm 
\epsffile{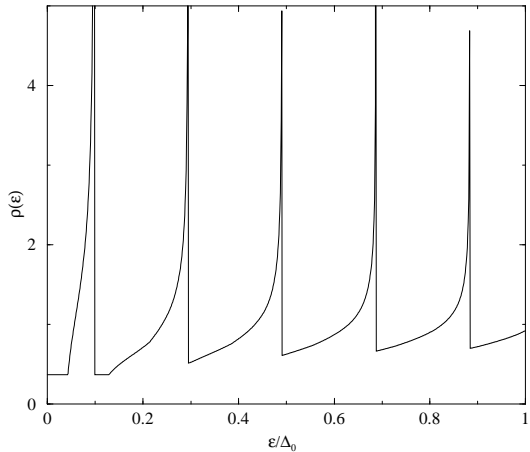}}}
\caption{The normalized DOS  (see text) in Bohr-Sommerfeld 
approximation as a function of $\varepsilon/\Delta_0$ 
for the disk geometry. 
The parameters are the same as in Fig.\ \ref{disk-raw}. With these
parameters $s_{\rm min}/\xi_0 = 15.0$. 
\label{abra-disk-rho-BS}}
\end{figure}

Thus, the total DOS for the NS disk system in Bohr-Sommerfeld approximation
is  $\rho_{\rm wgs}+\rho_{\rm AD}(\varepsilon)$.  
Since $P(s)$ is a singular function, the DOS becomes singular at the 
energies given by (\ref{ep-sing}) with 
$s_{\rm sing}= 2(R_{\rm N}-R_{\rm S})$.
In Fig.\ \ref{abra-disk-rho-BS} we plotted the normalized DOS, 
$\rho_{\rm BS}(\varepsilon)/(2\rho^{({\rm N})})$ in Bohr-Sommerfeld
approximation, where $\rho^{({\rm N})}=2m/\hbar^2(A/4\pi)$ is the DOS
of the normal circular annular billiard of area 
$A=(R_{\rm N}^2-R_{\rm S}^2)\pi$.
One can see that the DOS does not tend to zero as 
$\varepsilon \rightarrow 0$, contrary to the box geometry case. 
The non-vanishing DOS is due to the constant contribution of the whispering
gallery states given by Eq.\ (\ref{rho-wgs}).

\begin{figure}[hbt]
{\centerline{\leavevmode \epsfxsize=7cm 
\epsffile{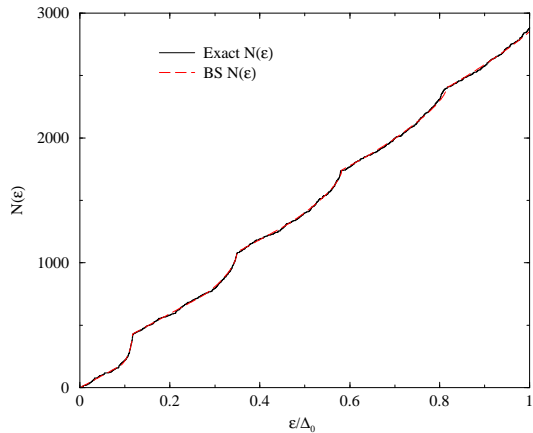}}}
\caption{The exact counting function $N(\varepsilon)$ (solid line) 
and $N_{\rm BS}(\varepsilon)$ given in (\ref{Nep-BS}) (dashed line) 
as functions of $\varepsilon/\Delta_0$ for the disk geometry. 
The parameters are the same as in Fig.\ \ref{disk-Weyl-1}. Then $s_{\rm
min}/\xi_0 =12.5$.
\label{disk-BS-2-A}}
\end{figure}

\begin{figure}[hbt]
{\centerline{\leavevmode \epsfxsize=7cm 
\epsffile{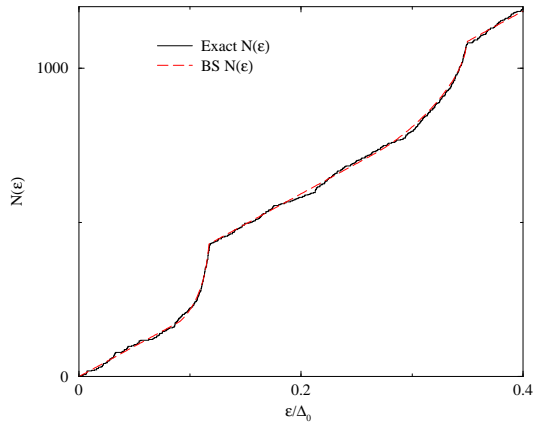}}}
\caption{Enlarged portion of Fig.\ \ref{disk-BS-2-A}. 
\label{disk-BS-2-B}}
\end{figure}

Again, one can determine the counting function $N_{\rm BS}(\varepsilon)$
by integrating the DOS.  
Using Eqs.\ (\ref{BS-Ne}) and (\ref{Ps-disk}) the integral 
can be performed analytically, and  
\begin{equation}
N_{\rm BS}(\varepsilon) = \rho_{\rm wgs}\, \varepsilon + 
2 k_{\rm F}R_{\rm S}\sum_{n=0}^\infty f(s_n(\varepsilon))\, 
\Theta (s_{\rm max}-s_n(\varepsilon))
\label{Nep-BS}
\end{equation}
is obtained, where
\begin{equation}
 f(s) = \left\{ \begin{array}{ll}
 1,  & \mbox{if}\,\,\,\, s \le s_{\rm min},  \\
 1-\frac{
\sqrt{\left[s_{\rm max}^4/s_{\rm min}^2 - s^2 \right]
\left[s^2 - s_{\rm min}^2 \right]}}{4\, s \, R_{\rm S}}
,& \mbox{if} \,\,\,\, s_{\rm min} < s \le s_{\rm max},
\end{array}   
 \right.  
\end{equation}
and $s_n(\varepsilon)$ is given by (\ref{sn-eps-rep}). Note that the
infinite sum over $n$ in (\ref{Nep-BS}) is indeed a finite one due to 
the Heaviside  function.
Using Eqs.\  (\ref{DNS})-(\ref{De}) we calculated the exact energy levels
for the NS disk system. In Fig.\ \ref{disk-BS-2-A} the exact counting 
function $N(\varepsilon)$ and $N_{\rm BS}(\varepsilon)$ are plotted 
for perfect interface (no mismatch and  zero tunnel barrier). 
One can see that the agreement between the exact counting function and
that obtained from the Bohr-Sommerfeld approximation is excellent for all
energies below the gap. 
To see the agreement, the details of Fig.\  \ref{disk-BS-2-A} -- 
up to the first two cusps -- are enlarged in Fig.\  \ref{disk-BS-2-B}.
Similarly, very good agreements were found for other parameter ranges of the
NS disk systems with perfect interface. 
 
In Fig.\ \ref{disk-raw} one can see that the energy levels 
(for each radial quantum number $n$) as functions of $m$ are 
approximately constant as $m\rightarrow 0$. 
Thus, the DOS should diverge because it is proportional to the reciprocal of 
the derivative of $\varepsilon_{mn}$ with respect to $m$ 
(assuming that $m$ is a continuous variable). 
The positions of the singularities of the DOS coincide with the energies 
$\varepsilon_{mn}$ at $m=0$ for all possible $n$. 
The angular momentum quantum number $m=0$ semiclassically 
corresponds to the radial orbits of the electron between the inner and 
outer circles. 
Applying the Bohr-Sommerfeld quantization rule for the electrons and the
Andreev reflected holes along the radius one finds
\begin{equation}
\left[k_e(\varepsilon) -k_h(\varepsilon)\right] (R_{\rm{N}}-R_{\rm{S}}) = 
n \pi +\arccos (\varepsilon/\Delta_0).
\label{BS-radial-eq}
\end{equation} 
Note that this equation can also be derived from the quantization condition 
given in Eq.\ (\ref{Phi_m}).
The solutions of Eq.\ (\ref{BS-radial-eq}) agree very well with the energies 
in Fig.\ \ref{disk-raw} at $m=0$. Therefore, the singularities of the DOS 
are related to the classical orbits along the radial directions. 

\begin{figure}[hbt]
{\centerline{\leavevmode \epsfxsize=7cm 
\epsffile{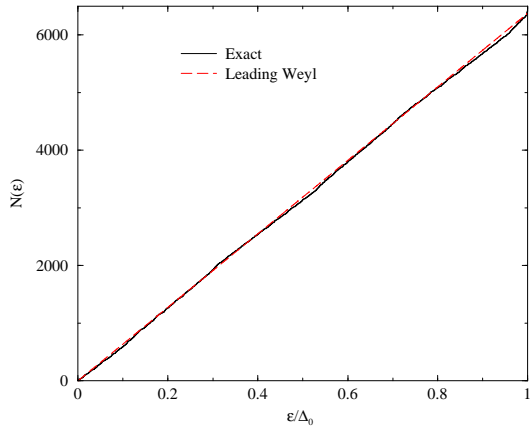}}}
\caption{The exact counting function $N(\varepsilon)$ (solid line) 
and the leading term of the Weyl formula $\tilde{N}(\varepsilon)$ 
(dashed line) as functions of $\varepsilon/\Delta_0$ 
for the NS disk geometry when the interface is not perfect. 
The parameters are $ R_{\rm S}/R_{\rm N} = 7/10$, 
$k_{\rm{F}}^{\rm{(N)}}R_{\rm S} = 350$,   
$\Delta_0/E_{\rm{F}}^{\rm{(N)}}=0.1$, 
$r_k=0.01$, $r_v=0.1$, and tunnel barrier, $Z=0$. 
\label{abra-mismatch-disk}}
\end{figure}

We also calculated the energy levels in the case of non-perfect NS
interface.  
Solving the secular equation (\ref{DNS}) numerically, the counting function 
$N(\varepsilon)$ is shown in Fig.\ \ref{abra-mismatch-disk}.
It can be seen from the figure that for non-perfect NS interface 
the cusps in the counting function are 'washed out' similarly to the
case of NS box systems.  This is related 
to the fact that the normal reflections at the interface are enhanced. 
Our numerical results show that the cusps also disappear for $Z\neq 0$.   
Using the Weyl formula derived in Sec.\ \ref{Weyl-disk} we found that 
$\tilde{N}(\varepsilon) = 2 \rho^{\rm{(N)}} \varepsilon$ in leading order, 
where $\rho^{\rm{(N)}}=
(2m/\hbar^2)\, \left(R_{\rm N}^2 - R_{\rm S}^2\right)/4$  is 
the density of states for normal annular billiard with inner radius 
$R_{\rm S}$ and outer radius $R_{\rm N}$. 
In Fig.\ \ref{abra-mismatch-disk} this is drawn
by a dashed line. One can see that this Weyl formula is a good approximation
for the exact counting function in the case of non-perfect interfaces. 
The result shows that the role of normal reflections dominates over 
Andreev reflections, and thus the Weyl formula agrees approximately with
that for the circular annular billiard without 
superconducting region.

\section{`Phase diagram' for NS disk systems}
\label{fazis-diag}

In our studies of the energy levels of the NS disk systems we have
assumed so far that the energy dependent coherence length 
$\xi(\varepsilon)=\hbar v_{\rm F}/\sqrt{\Delta_0^2-\varepsilon^2}$ 
is much smaller than $R_{\rm S}$ (see the text after 
Eq.\ (\ref{coherence-xi})). 
In this section we shall discuss the case when 
$\xi(\varepsilon) \gg R_{\rm S}$, i.e.\  when 
$\varepsilon \rightarrow \Delta_0$. The imaginary part of  
$q_e R_{\rm S} \approx k_{\rm{F}} R_{\rm S} + i\, R_{\rm
S}/\xi(\varepsilon)$ is then negligible, and 
$k_e  R_{\rm S}\approx q_e R_{\rm S}$. 
Thus, in the determinant $D^{(e)}_m(\varepsilon)$ given 
in Eq.\ (\ref{De-disk}) the arguments of the Bessel functions are 
almost equal. We now consider the perfect NS interface 
($Z=0$ and $m_{\rm N}=m_{\rm S}$). 
Subtracting the second column from the first one in the determinant, 
the ratio  
$J_m(k_e R_{\rm N})/Y_m( k_{e}R_{\rm N})$ can be factored out. 
It turns out that the remaining determinant is non-zero 
for $\varepsilon < \Delta_0$. 
Thus, similarly to the whispering gallery states discussed 
in Sec.\ \ref{Weyl-disk}, it holds to a very good approximation that 
$D^{\rm{(e)}}_m(\varepsilon)$ (as a function of $\varepsilon$) 
has zeros whenever  $J_m(k_e R_{\rm N})=0$ for {\it all} $m$. 
This corresponds to the energy
levels of a circular billiard of radius $R_{\rm N}$ for electron-like
quasiparticles. The same is true for the hole-like quasiparticles, 
i.e.\ their energy levels are the solutions of 
$J_m(k_h R_{\rm N})=0$ for all $m$. 
In summary, we found that for $\xi(\varepsilon)/R_{\rm S}\gg 1$ the NS
disk system behaves as a normal full disk, while 
in the opposite case 
the energy levels of the system have a `mixed phase' character, in which 
Andreev states and whispering gallery states coexist 
(see Sec.\ \ref{Weyl-disk}). 
We shall call the first case a `full disk phase' (FD). 
According to our numerical results, it is safe to say that 
the MP and FD can be distinguished very well by the ratio  
$\xi(\varepsilon)/R_{\rm S}$. 
We found mixed phase for  $\xi(\varepsilon)/R_{\rm S}< 2$ and 
FD phase for $\xi(\varepsilon)/R_{\rm S}> 2$.  

As it is known, the Andreev approximation is valid only for
$\Delta_0/E_{\rm F} \ll 1$. If we assume some critical value, say  
$\Delta_0/E_{\rm F} = 0.1$, 
then for $k_{\rm F}\xi_0 =2 E_{\rm F}/\Delta_0 < 20$ 
the Andreev approximation is not valid and normal reflection 
at the NS interface dominates over Andreev reflection. 
The system looks like an annular billiard 
(without superconductor region). 
We shall call this case an `annular phase' (AP). 
As we have seen in the Fig.\ \ref{abra-mismatch-disk}, the 
non-perfect interface also results in an enhanced normal reflection 
and the character of the energy levels again corresponds 
to a circular annular billiard (annular phase). 

\begin{figure}[hbt]
{\centerline{\leavevmode \epsfxsize=7cm 
\epsffile{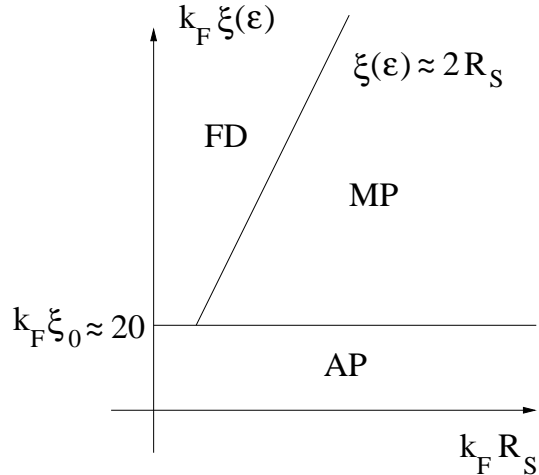}}}
\caption{Schematic  phase diagram for the NS disk systems.
The mixed phase (MP) is bounded by the lines $k_{\rm F}\xi_0 \approx 20$ and 
$\xi(\varepsilon)\approx 2R_{\rm S}$. 
The full disk phase (FD) is bounded by  
the line $k_{\rm F}R_{\rm S} =0$, the line $k_{\rm F}\xi_0 \approx 20$
and the line $\xi(\varepsilon)\approx 2R_{\rm S}$. 
The annular phase (AP) is below the line $k_{\rm F}\xi_0 \approx 20$. 
\label{abra-fazis}}
\end{figure}

We calculated the exact energy levels for different parameters and 
compared with those obtained for the full disk and for circular 
annular disk. In this way, we can classify the NS disk
systems into the MP, FD and AP phase regions.
The crossovers between the three `phases' are not sharp and 
depend on the energy, too.
It turns out that the three phases can be characterized by 
two parameters, 
$k_{\rm F} R_{\rm S}, k_{\rm F}\xi(\varepsilon)$.
These are the relevant parameters
for the NS disk systems and each pair corresponds to one of the phases.
Thus, a so-called phase diagram can be given for the different
phases. The boundaries of these phases are not so sharp, however. 
Far from the boundaries the given phase becomes dominating.  
In Fig.\ \ref{abra-fazis} the three different phases are shown.

\section{Conclusions} \label{veg}

We investigated the energy spectrum of NS box and disk systems 
using the BdG equation. Matching the wave functions at the NS interface 
we derived a secular equation whose solutions give the energy spectrum
of the system.
The mismatch in the Fermi wave numbers 
and the effective masses of the normal system and the superconductor, 
as well as the tunnel barrier is included in the calculation.
Rewriting the secular equation a simple quantization condition was
derived.
Equation (\ref{Phi_m}) is a central result and serves as a starting point 
for further theoretical studies in this paper.  
Using this quantization condition (a) we
derived for both NS systems with perfect interfaces a Weyl formula 
for the counting function,  
(b) by re-deriving the commonly used Bohr-Sommerfeld approximation 
to the DOS we presented an explicit expression for the probability
$P(s)$ in terms of the classical action of the electron moving in the N
region between two successive bounces at the NS interface. 
The numerically exact calculations are in very good agreement 
with our Weyl formula and the Bohr-Sommerfeld approximation. 
The singularities in the DOS obtained from our exact calculations can be 
successfully explained by using the correct probability function $P(s)$. 
$P(s)$ is a singular function of $s$ both for NS box and disk systems. 
According to our theory concerning the singularities, this implies that the
singularities of the DOS are located at equal distances from each
other.
A simple formula is given for the location of the singularities. 
We demonstrated that it can be applied for the system studied by de
Gennes et al.\ \cite{deGennes-Saint-James}, and it may as well give the
reason for the significant peaks observed by 
Ihra et al.\ \cite{Richter1} in their numerical calculations.
We show that the singularities in $P(s)$ are related to some special 
classical trajectories of the electron moving in the N region hitting
the NS interface.

In the case of NS disk systems the DOS is constant as the energy 
gets close to the Fermi level. This is, at first sight, in contrast 
to the common belief that for integrable billiards the DOS should go  
to zero as the energy tends the Fermi level. We pointed out that 
in this NS disk system the whispering gallery states give the 
non-vanishing DOS at the Fermi level. 
However, the Andreev states still have no 
contribution to the DOS. In the disk system the Andreev and the
whispering gallery states coexist (so-called mixed phase). 
Depending on the geometry of the disk and
the material parameters, we sketch a kind of phase diagram to
classify the energy spectra into three classes: (i) mixed phase, (ii)
full disk, (iii) annular disk. If the coherence length is much larger
than the diameter of the superconducting region, the spectrum
corresponds to the full disk phase. The electrons travel thorough the S
region without Andreev reflection. When the Andreev approximation
fails, the normal reflection is enhanced at the NS interface and the
system behaves as an annular circular billiard. This is also the case
when there is a tunnel barrier at the NS interface or 
a mismatch between the effective masses and the Fermi wave numbers 
of the normal and superconducting regions.

An interesting extension of the study presented in this paper may be
the investigation of the role that the geometry of the normal billiard plays 
in the singularities of the DOS. Under what conditions do such
singularities exist?
Does the energy spectrum at higher energies show some special structure 
in chaotic billiards? How can the Weyl formula (derived for NS box
and disk systems) be extended for other integrable and chaotic
billiards?
Our quantization method can also be applied for NS disk in the presence of 
a magnetic field.  In this case, besides Andreev states and whispering
gallery states (more specifically edge states) 
one has to take into account the Landau levels that appear 
in the spectrum in the presence of strong magnetic fields.
 These  give rise to a much more complex energy spectrum than 
a zero magnetic field. 
Another relevant question is the study of 
the counting function in these systems. Further work along these lines 
is in progress.

\acknowledgements 

One of us (J.\ Cs.) would like to thank C. W. J. Beenakker, U.\ Z\"ulicke, 
T.\ Geszti, Z.\ Kaufmann and A. Pir{\'o}th for useful discussions. 
This work supported in part by the European Community's Human Potential
Programme under contract HPRN-CT-2000-00144, Nanoscale Dynamics, and 
the Hungarian Science Foundation (OTKA TO25866 and TO34832).

\appendix 
\section{Calculation of the phase 
$\Phi_{\protect\lowercase{m}}(\varepsilon)$}
\label{app-disk}

It is clear for symmetry reasons that the eigenstates with quantum 
numbers $m$ and $-m$ are degenerate (except for $m=0$),  
thus it is enough to consider the states $m \ge 0$. 
In the Andreev approximation 
($\Delta_0/E_{\rm{F}} \ll 1$) we take 
$k^{(e)}_m \approx q^{(e)}_m$ in places where they are multiplied 
by the Bessel functions in $D^{\rm{(e)}}_m(\varepsilon)$ but 
in the arguments of the Bessel functions we keep them to be different.
As we shall see, different methods are necessary 
to approximate the determinant $D^{\rm{(e)}}_m(\varepsilon)$ 
for different ranges of $m$.   
The Debye asymptotic expansion\cite{Abramowitz} will be used to approximate
the determinant for $|m| < k_e R_{\rm S}-\sqrt[3]{k_e R_{\rm S}}$.
In this expansion the Bessel functions with the real argument 
$x$ for $0< m < x-\sqrt[3]{x}$ are approximated by
\begin{eqnarray}
J_m(x) &\approx& \sqrt{\frac{2}{\pi}} \frac{1}{\sqrt{x\sin Q}} \,
\cos \left[x\left(\sin Q -Q \cos Q\right) -\frac{\pi}{4}\right], 
\label{Bessel-Debye} \\[1ex]
Y_m(x) &\approx& \sqrt{\frac{2}{\pi}} \frac{1}{\sqrt{x\sin Q}} \,
\sin \left[x\left(\sin Q -Q \cos Q\right) -\frac{\pi}{4}\right], 
\end{eqnarray}
where $\cos Q= m/x$. 
For $|m| > k_e R_{\rm S}+\sqrt[3]{k_e R_{\rm S}}$ 
an analogous expansion\cite{Abramowitz} can be used. 
In the intermediate range  $|m-k_e R_{\rm S}| < \sqrt[3]{k_e R_{\rm S}}$ 
one may apply uniform asymptotic expansions\cite{Abramowitz} 
for the Bessel functions.   
For  Bessel functions  of a complex argument $z$  with fixed $m$  
the approximate form\cite{Abramowitz} 
\begin{equation}
J_m(z) \approx \sqrt{\frac{2}{\pi z}} \cos \left( z-\frac{1}{2}m \pi 
-\frac{\pi}{4} \right), \,\,\,\,\, {\rm for}\,\,\,\,\, z\rightarrow \infty 
\label{complex-Bessel}
\end{equation}
will be used. 
We first calculate the eigenphase $\Phi_m (\varepsilon)$ for 
$|m| < k_e R_{\rm S}-\sqrt[3]{k_e R_{\rm S}}$.  
Using Eqs.\ (\ref{Bessel-Debye})-(\ref{complex-Bessel}) for the Bessel 
functions  in Eqs.\  (\ref{De-disk}) and (\ref{S_NS}), one finds
\begin{equation}
e^{i \Phi_m(\varepsilon)} =
\frac{1+e^{-2i \left[ \vartheta_m(k_e R_{\rm N}) - \vartheta_m(k_e R_{\rm S}) 
+\beta^*\right]}}{
1+e^{2i \left[ \vartheta_m(k_e R_{\rm N}) - \vartheta_m(k_eR_{\rm S}) 
+\beta \right]}}
\, e^{2i\left[\vartheta_m(k_e R_{\rm N}) 
-\vartheta_m(k_eR_{\rm S}) \right]} \, e^{i\left(\beta +\beta^*\right)},
\label{phi-disk}
\end{equation}
where 
\begin{eqnarray}
\vartheta_m(x)  & = & \sqrt{x^2-m^2}
-\left|m\right| \arccos \frac{\left|m\right|}{x}, 
\label{eta-def} \\
\beta & =  & q_e R_{\rm S}-\frac{1}{2}m\pi -\frac{\pi}{4}.
\end{eqnarray}
The absolute values of $m$ in the above expressions are the consequence of
degeneracy of the eigenstates $m$ and $-m$.   
Note that $\beta$ is a complex number.  
Thus, for $\Delta_0 \ll E_{\rm F}$ and below the gap 
$\eta \ll 1$,  Eq.\ (\ref{qe-disk}) yields 
\begin{equation}
\beta = 
k_{\rm{F}} R_{\rm S}-\frac{1}{2}m \pi -\frac{\pi}{4} 
+ i\, \frac{R_{\rm S}}{\xi(\varepsilon)},
\end{equation}
where the coherence length $\xi(\varepsilon)$ is defined\cite{BdG-eq} as  
\begin{equation}
\xi(\varepsilon)= 
\frac{\hbar v_{\rm F}}{\sqrt{\Delta_0^2 - \varepsilon^2}}  
= \frac{2}{k_{\rm{F}}|\eta|}.
\label{coherence-xi}
\end{equation}
Here $v_{\rm F}$ is the Fermi velocity.
We now assume that  $R_{\rm S} \gg \xi(\varepsilon)$. 
The opposite case will be discussed in Sec.\ \ref{fazis-diag}.   
Then, the first factor in Eq.\ (\ref{phi-disk}) is 
approximately equal to 1, and we obtain
 \begin{equation}
\Phi_m(\varepsilon) = 
2\left[\vartheta_m(k_e R_{\rm N}) -\vartheta_m(k_eR_{\rm S}) \right]
+ 2k_{\rm{F}} R_{\rm S} -m \pi -\frac{\pi}{2}. 
\label{phi-disk-1}
\end{equation}

We now turn to the case 
$|m| > k_e R_{\rm S}+\sqrt[3]{k_e R_{\rm S}}$.  
In this case, according to the analogous Debye asymptotic expansion of 
the Bessel functions, one finds that $J_m(k_e R_{\rm S})$ 
can be neglected (it is exponentially small) 
compared to the second term of the 1,1 element of the determinant 
in (\ref{De-disk}). Similarly, 
$J_m^\prime(k_e R_{\rm S})$ is also negligible
in the determinant. Thus, the ratio  
$J_m(k_e R_{\rm N})/Y_m( k_{e}R_{\rm N})$ can be taken out from the
determinant. It turns out that the remaining determinant is not equal to 
zero for $\varepsilon < \Delta_0$. 
Thus, to a very good approximation,  
$D^{\rm{(e)}}_m(\varepsilon)$ (as a function of $\varepsilon$) 
has zeros whenever  $J_m(k_e R_{\rm N})=0$. This corresponds to the energy
levels of a circular billiard of radius $R_{\rm N}$ for electron-like
quasiparticles. Note that now $\Phi_m(\varepsilon)$ cannot be calculated  
from Eq.\ (\ref{S_NS}), since $D^{\rm{(e)}}_m(\varepsilon)\approx 0$ for 
$|m| > k_e R_{\rm S}+\sqrt[3]{k_e R_{\rm S}}$. 
In this case we shall use a different method to determine the 
contributions to the smooth part of the counting function.

To approximate $D^{\rm{(e)}}_m(\varepsilon)$ in the intermediate range 
$|m-k_e R_{\rm S}| < \sqrt[3]{k_e R_{\rm S}}$, one may apply  uniform
asymptotic expansions for Bessel functions. Semiclassically, these
states are related to the diffraction of the electron in the penumbra of 
a circular annulus billiard\cite{Primack_Snaith}. As a first estimate of 
the smooth part of the counting function, the energy levels in the 
intermediate range will be taken into account by the corresponding energy
levels of the circular billiard of radius $R_{\rm N}$ using 
Debye's asymptotic expansion. 
As we shall see, this is a good approximation since the intermediate range 
of $m$ is rather narrow compared to the two other ranges discussed above. 
Using the uniform expansions of the Bessel functions 
in our derivation of the Weyl formula for the NS disk system, 
it is straightforward (although algebraically rather tedious) 
to include the intermediate range more accurately.

Similarly, the method of approximations outlined above 
can be used to approximate $D^{\rm{(h)}}_m(\varepsilon)$ 
in the secular equation (\ref{DNS}).
One finds that  $\Phi_m(-\varepsilon)$ is still  
given by (\ref{phi-disk-1}) after replacing $k_e$ by the wave number 
$k_h(\varepsilon) =k_e(-\varepsilon)$ of the hole-like quasiparticles in
the N region. However, in this case the approximation for  
$\Phi_m(-\varepsilon)$ is valid only for 
$|m| < k_h R_{\rm S}-\sqrt[3]{k_h R_{\rm S}}$.
Finally, from Eq.\ (\ref{phi-disk-1}) one obtains 
\begin{equation}
\Phi_m (\varepsilon)-\Phi_m (-\varepsilon) = 
2\left[\vartheta_m(k_e R_{\rm N}) -\vartheta_m(k_e R_{\rm S}) \right]
-2\left[\vartheta_m(k_h R_{\rm N}) -\vartheta_m(k_h R_{\rm S}) \right]
\label{Fi-kul}
\end{equation}
for 
$|m| < k_h R_{\rm S}-\sqrt[3]{k_h R_{\rm S}}$.
Note that the last three terms in (\ref{phi-disk-1}) cancel out in
the difference $\Phi_m (\varepsilon)-\Phi_m (-\varepsilon)$.

In a similar way, it can be seen that the zeros of 
$D^{\rm{(h)}}_m(\varepsilon)$ in Eq.\ (\ref{DNS}) coincide with the
zeros of $J_m(k_h R_{\rm N})$ 
for $|m|> k_h R_{\rm S}+\sqrt[3]{k_h R_{\rm S}}$ to a good approximation. 
Thus, the secular equation
(\ref {DNS}), determining the energy levels of the NS disk system, 
reduces to  
\begin{equation}
J_m(k_e R_{\rm N})J_m(k_h R_{\rm N})=0 
\label{whisp-sec}
\end{equation}
for  $|m| > k_e R_{\rm S}+\sqrt[3]{k_e R_{\rm S}}$. 
The energy levels are the same as those of the disk of radius $R_{\rm N}$
for electron/hole-like quasiparticles. 
Semiclassically, the states with angular momentum quantum number 
$|m|> k_e R_{\rm S}+\sqrt[3]{k_e R_{\rm S}}$ 
are called whispering gallery modes. 
The wave functions for these
states at $r=R_{\rm S}$ are negligible, and so no Andreev reflections take
place. This is why the superconducting pair potential
$\Delta_0$ does not appear in (\ref{whisp-sec}).

\end{document}